\documentclass[a4paper,fleqn,usenatbib]{mnras}

\usepackage{newtxtext,newtxmath}
\usepackage[T1]{fontenc}
\usepackage{ae,aecompl}

\usepackage[dvipdfmx]{graphicx}	% Including figure files
\usepackage{amsmath}	% Advanced maths commands
\usepackage{amssymb}	% Extra maths symbols

\usepackage{wasysym,verbatim,lscape,rotating,hyperref}
\usepackage{multirow}
\usepackage{float}
\usepackage{color}
\usepackage{dblfloatfix}

\newcommand{\bes}{Besan\c{c}on}
\newcommand{\euclid}{\emph{Euclid}}

\newcommand{\wfa}{\emph{WFIRST}}
\newcommand{\lsst}{LSST}
\newcommand{\moa}{MOA}
\newcommand{\ogle}{OGLE}
\newcommand{\kepler}{\emph{Kepler}}
\newcommand{\ts}{\textstyle}

\title[Probability of FFP parallax detection]{Probability of simultaneous parallax detection for free-floating planet microlensing events near Galactic Centre}

\author[Ban]{
M. Ban $^{1}$ %and A.C. Robin $^{2}$
\\
$^{1}$National Astronomical Observatory of Japan, Mitaka Campus, Tokyo, Japan\\
}

\date{Accepted XXX. Received YYY; in original form ZZZ}

\pubyear{2020}

\hypersetup{draft}		% to avoid fatal error due to the separation of hyperlink
\begin{document}
\label{firstpage}
\pagerange{\pageref{firstpage}--\pageref{lastpage}}
\maketitle

% Abstract of the paper
\begin{abstract}
The event rate and the efficiency of mass estimation for free-floating planet (FFP) microlensing events were determined from the simulation of the simultaneous parallax observations by \euclid{}, \wfa{}, and \lsst{}. The stellar population from the \bes{} Galactic model toward $(l, b)=(1^{\circ},-1.^{\circ}75)$ was applied to our 3D microlensing model, and 30,000 parallax observations were simulated for each following FFP lens masses: Jupiter-mass, Neptune-mass, and Earth-mass assuming the population of one FFP per star. The interstellar dust, unresolved stellar background, nearby star blending were modelled. A signal-to-noise limit considering a finite source effect determined the event detectability. The \euclid{}-\wfa{} combination yielded 30.7 Jupiter-mass FFPs during two 30-day-periods per year in parallax observation. The parallax event rate decreases to 3.9 FFPs for Earth-mass planets. The mass estimation from the parallax light curve allowed recovery of FFP masses to within a factor of two for 20-26\% of cases. The \euclid{}-\lsst{} combination yielded 34.5 Jupiter-mass FFPs down to 0.5 Earth-mass FFPs for the same periods and the mass is recovered to within a factor of two in 20-40\% of cases. The event rate will be normalised by the unknown FFP abundance to recover the number of expected detections.
\end{abstract}

% Select between one and six entries from the list of approved keywords.
% Don't make up new ones.

%%%%%%%%%%%%%%%%%%%%%%%%%%%%%%%%%%%%%%%%%%%%%%%%%%

%%%%%%%%%%%%%%%%% BODY OF PAPER %%%%%%%%%%%%%%%%%%

\section{Introduction}		\label{sec:intro}
Microlensing event observation is a useful method for exoplanet research. Some wide-orbit bound planets and candidate free-floating planets (FFPs) have been found in recent decades (\citealt{Beaulieu2006, Gould2006, Muraki2011, Sumi2011, Sumi2013, Mroz2018}). A single observation of microlensing event enables calculation of the lens size (so-called Einstein radius) from the magnification and event duration, but it is still a challenge to identify the details of the lens properties. A microlensing parallax observation is, therefore, expected to yield additional keys for analysis of the lens properties. Stereo-vision offers different sightlines towards a microlensing event, and the lens mass and distance can be calculated more effectively.

Parallax observation has long been used in the history of astronomy based on the Earth's position with respect to the Sun. For exoplanet research, however, the caustic magnification of a microlensing effect by a planetary object is much quicker. Therefore, simultaneous parallax observation by separated observers is a more appropriate method. \citet{Poindexter2005} analysed some microlensing events with ground-based parallax observation data and found a candidate Jupiter-mass FFP. \citet{Mogavero2016} suggested the efficiency of ground-based and geosynchronous satellite parallax for FFP search, and recently, the parallax observation of ground-based and space-based telescopes is developing. For example, the space-based telescope \kepler{} conducted microlensing parallax observations with the ground-based telescopes such as MOA, OGLE, and other >25 telescopes \citep{Henderson2016a, Henderson2016b, Gould2013, Zhu2017}. The data observed by {\it Spitzer} were applied to analyse OGLE data, and some planetary objects were reported (\citealt{Zhu2015, Novati2015, Street2016}). Moreover, some new telescope missions have been proposed and are expected to be operational within recent 10 years. Especially, \euclid{} and \wfa{} are expected to find exoplanets including FFPs through microlensing events and parallax observations \citep{Penny2013, Mcdonald2014, Hamolli2014, Zhu2016}.

\begin{figure*}	
\centering
\begin{tabular}{cc}
\hspace{-0.1in}\includegraphics[width=3.5in]{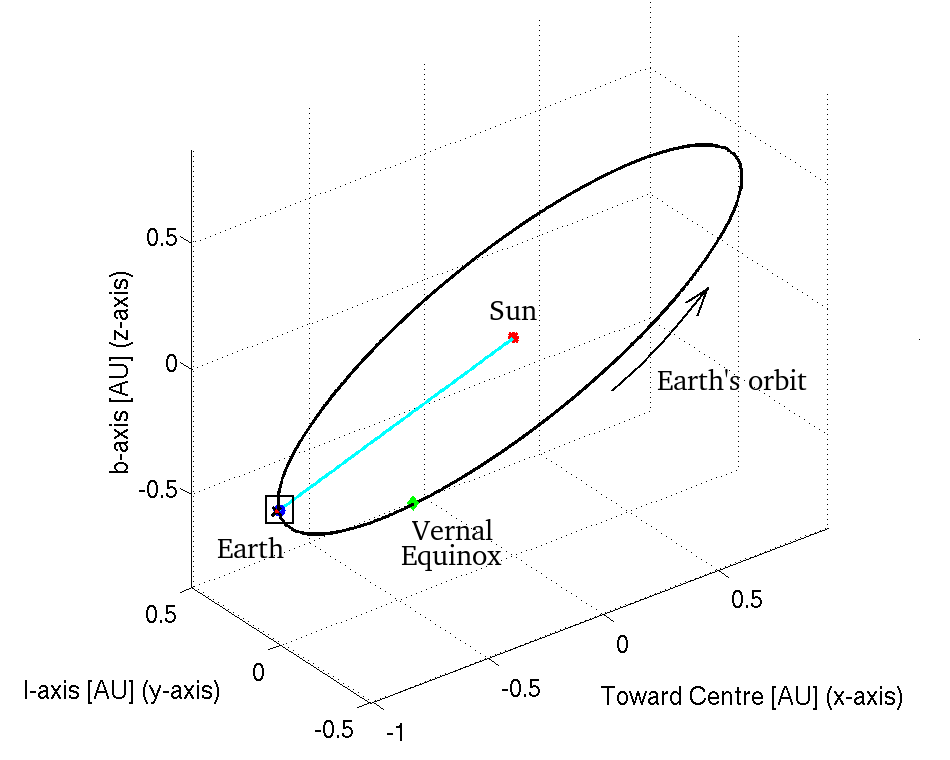}	&
\hspace{-0.1in}\includegraphics[width=3.5in]{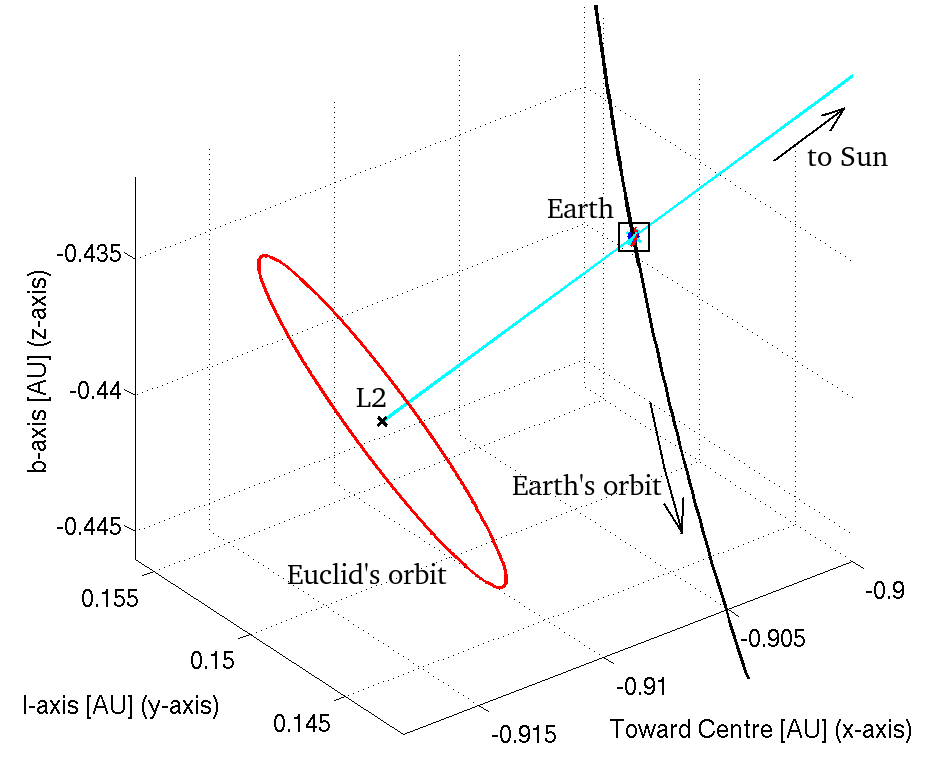} \\
(a) Earth motion & (b) \euclid{} motion\\
\end{tabular}
\caption{(a) 3D model of the Earth (i.e. \lsst{}) in the Galactic Coordinate System (GCS). The longitude and latitude are along the y-axis and z-axis, respectively. The x-axis shows the direction toward the Galactic centre. The small square around the Earth is the zoom-up window for (b). (b) 3D model of the \euclid{} orbit in the GCS. The circle around L2 is in the terrestrial sky frame so that it always perpendicular to the Sun-Earth-L2 line. We assume \wfa{} shares this orbit in case of the Halo orbit at L2. } 
\label{fig:3d}
\end{figure*}

In this paper, we report the simulation of parallax observation of FFP microlensing events in a 3D model with signal-to-noise consideration to derive the effective value of parallax event rate. \citet{Henderson2016b} also considered a simultaneous parallax observation of FFP microlensing for the combination of \kepler{} and some ground-based surveys. Because of \kepler{}'s motion, the separation between these two telescopes varies; and they obtained the result that Earth-mass FFPs could be detected in parallax at the early stage of K2 Campaign 9. They mentioned the importance of observer separation with respect to the lens size. Hence, we chose the target area of $(l, b)=(1^{\circ}, -1.^{\circ}75)$ monitoring by two combinations of separated telescopes: \euclid{}-\wfa{} and \euclid{}-\lsst{}. Their separation is not as variable as \citeauthor{Henderson2016b}'s combination and the shorter baseline allows more effective targetting of low-mass FFPs due to their smaller Einstein rings. In \S\ref{sec:tele}, the configuration of three telescopes (\euclid{}, \wfa{} and \lsst{}) is reviewed. In \S\ref{sec:micro}, we describe the simulation process of FFP microlensing events using \bes{} model data without considering parallax observation. In \S\ref{sec:prlx}, the processes of event initialisation and parallax simulation are explained. The results will be shown in \S\ref{sec:res} and discussed in \S\ref{sec:disc}, and conclusions are described in \S\ref{sec:conc}.

\section{Observatory configuration}		\label{sec:tele}
We assume \euclid{} is the main telescope in our simulation, and \wfa{} and \lsst{} will be partners for simultaneous parallax observation of FFP events. Hence, the combination of \euclid{} and \wfa{} offers the parallax in similar sensitivity (i.e. space-based $H$-band), and the combination of \euclid{} and \lsst{} offers the parallax in different photometric bands.

\subsection{Conditions}	\label{subsec:tele-kine}
\euclid{} is expected to launch in 2022\footnote{As of 25th February 2020 retrieved from \euclid{} mission website \url{http://sci.esa.int/euclid/}} and orbit around Lagrange Point 2 (L2) with a period of 6 months \citep{ESA2011}.  In the terrestrial sky, the angular distance of \euclid{} from L2 is no more than 33 deg, and the solar aspect angle (SAA) must be kept within 90$^{\circ}$<SAA<120$^{\circ}$. It limits the possible observation period to $\sim$30 days around the equinoxes.

The Wide Field Infrared Survey Telescope (\wfa{}) will be launched sometime around 2024 \citep{Spergel2015}. The geosynchronous orbit around the Earth with a distance of 40,000 km and the Halo orbit around L2 were considered in the planning stage, and the Halo orbit was decided. In our simulation, both orbits are taken and will be compared. In case of the geosynchronous orbit, the orbital path inclines 28.5 deg from the celestial equator and a node located at RA=175 deg. The distance and inclination avoid occultations by the Earth when targeting at the hot spot. In case of the Halo orbit at L2, some trajectories have been discussed and is expected to be similar to the \euclid{} trajectory \citep{Folta2016, Bosanac2018}. We assume it shares the \euclid{} orbital period with the orbital radius of $\sim0.75\times10^6$ km \citep{Webster2017} in our simulation. \wfa{} allows wider SAA as 54$^{\circ}$<SAA<126$^{\circ}$ which completely covers the \euclid{} observation period. Thus, \euclid{} will determine the observation period of simultaneous parallax detection for the \euclid{}-\wfa{} combination.

The Large Synoptic Survey Telescope (\lsst{}) is an 8m-class telescope being built at Cerro Pach\'{o}n in Chil\'{e} which will start operations in 2022 \citep{Ivezic2008}.  It has not scheduled the high cadence, continuous operation for the microlensing event in the campaign as of today. Nonetheless, we selected this telescope to exemplify ground-based surveys. The sensitivity has enough potential to operate microlensing observations for FFPs down to the Earth-mass size whilst the current microlensing missions such as \moa{} and \ogle{} are relatively difficult to detect such low-mass lens events. Unlike the space-based telescopes described above, the day-night time and airmass will limit the observation period. Targeting at the Galactic centre, we assume \lsst{} can on average perform for 7.5 hours per night\footnote{This value is derived from \url{http://www.eso.org/sci/observing/tools/calendar/airmass/html} which we assumed a site of La Silla Observatory is a close proxy location. The airmass 1$\leq $sec(z)$\leq$8 was taken.} during the \euclid{} observation period. Moreover, we assume the fine weather for photometric observation is $\sim$65.89\% by taking an average of the climate data from 1991 to 1999 at La Silla observatory.\footnote{\url{http://www.eso.org/sci/facilities/lasilla/astclim/weather/tablemwr.html}} We predict that, even with the sensitivity of \lsst{}, we cannot cover the event detections as reliably as the space-based surveys. Hence, we simulate the \euclid{}-\lsst{} combination as an indication of the ground-based sensitivity limitation and for comparison with the \euclid{}-\wfa{} combination.

Figure \ref{fig:3d} shows a 3D image of the telescope motion in the Galactic Coordinate System (GCS). We assume that the \wfa{} phase is $90^{\circ}$ ahead of the \euclid{}phase in the x-y sky frame centred on L2. The barycentric motion of the Earth and Moon is ignored since the barycentre exists within the Earth and our Monte-Carlo based simulation (the detail of sample data pick-up is explained later) moderates the uncertainty due to the barycentric motion. An appropriate distance between two telescopes is an important factor of simultaneous parallax observation. We expect the combinations of \euclid{}-\wfa{} and of \euclid{}-\lsst{} would show a geometrical factor on their parallax detectability.

\begin{table}
 \centering
 %\begin{minipage}{140mm}
  \caption{Survey parameters for three telescopes. The \euclid{} sensitivity is taken from Table 2 of \citet{Penny2013}, the \wfa{} sensitivity is from \citet{Spergel2015} and the \lsst{} sensitivity is from \citet{Ivezic2008,LSST2009}. We will handle sample source data in the Johnson-Cousins photometric system throughout our simulation, therefore a proxy band (in the row of ``J-C photometry'') is assumed for every telescope filter. }
\label{tab:survs}
  \begin{tabular}{@{}lp{0.7in}p{0.7in}p{0.7in}@{}}
  \hline
  & \lsst{} & \euclid{} &\wfa{}\\ \hline
  Location & ground-based & space-based & space-based \\
  Filter & $z$ & NIPS $H$ & $W149$ \\
  J-C photometry & $I$ & $H$ & $H$ \\  \hline
  $u_{\rm max}$ & 3 & 3 & 3\\
  $m_{\rm sky}$ [mag/arcsec$^2$] & 19.6 & 21.5 & 21.5 \\	
  $\theta_{\rm psf}$ [arcsec] & 1.3 & 0.4 & 0.4\\
  $m_{\rm zp}$ & 28.2 & 24.9 & 27.6 \\
  $t_{\rm exp}$ [sec] & 30 & 54 & 52 \\
  %$\eta_T$ & 0.80 & 0.90 & 0.92 \\
\end{tabular}
%\end{minipage}
\end{table}

\subsection{Parameters} \label{subsec:tele-param}

We apply the Near Infrared Spectrometer and Photometer (NISP) $H$-filter of \euclid{} and $W149$-filter of \wfa{} for a space-based survey in our simulation \citep{ESA2011, Spergel2015}. These filters have very similar transmission curves, so closely approximate each other. For comparison to the \lsst{} $z$-band filter \citep{Ivezic2008}, we use the approximation of the Johnson $I$-band filter. The source magnitude is treated with the Johnson-Cousins photometric system in our simulation; hence we approximate the bands to the filters. Table \ref{tab:survs} summarises the telescope parameters for FFP surveys. 
The formula for microlensing amplitude is defined as 
\begin{equation}	\label{eq:au}
A(t) = \frac{u(t)^2+2}{u(t)\sqrt{u(t)^2+4}},
\end{equation}
where $A(t)$ the amplitude of the detected flux and $u(t)$ is the impact parameter in units of Einstein radii. We assume \euclid{}, \wfa{} and \lsst{} are sensitive enough to start a microlensing observation when the lens is approaching the source with the projected distance of 3 times larger than the Einstein radius. Thus, the maximum impact parameter is set as $u_{\rm max}$=3 for the point source case. $u_{\rm max}$=3 corresponds to the minimum amplitude of $A_{min}$$\sim$1.02. For the finite source case, however, Eq.(\ref{eq:au}) is no longer sufficient.

\begin{figure}
\hspace{-0.1in}
\includegraphics[width=3.8in]{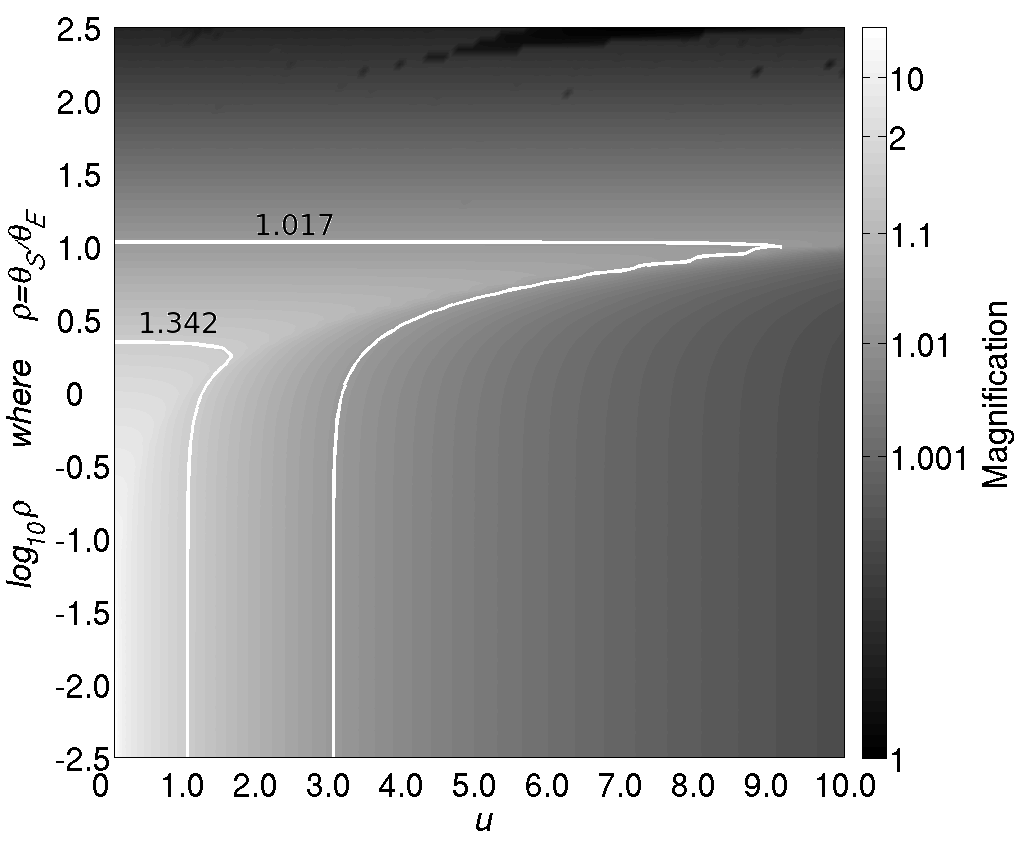}
\caption{Finite source magnification with impact parameters ($u$) on the x-axis and angular source radius in units of Einstein radius ($\log_{10}\rho$) on the y-axis. The two white contours correspond to two threshold amplitudes with $u_{\rm max}$=1 and $u_{\rm max}$=3 for the point source regime. This figure is reproduced from Fig. 1 in \citet{Ban2016}. }
\label{fig:amp}
\end{figure}

Figure \ref{fig:amp} shows the distribution of maximum magnification in the finite source case, along with different impact parameters and source surface angular sizes in units of Einstein radius ($\rho$) retrieved from our previous paper \citep{Ban2016}. The contour labelled $1.017$ is the amplitude limit that we are assuming for our parallax simulation (i.e. $A_{min}$$\sim$1.02) and the -0.5$\leq$$ log_{10}\rho$$\leq$1.0 regime shows the strong ``boost'' of threshold impact parameter to yield $A_{min}$$\sim$1.02. It implies that some specific events (i.e. finite source with -0.5$\leq$$log_{10}\rho$$\leq$1.0) allow the telescopes to observe them even though the minimum approach of the lens is larger than $u_{\rm max}$. In our simulation, this finite source effect is taken into account. The sky brightness ($m_{\rm sky}$) and the full width at half-maximum size of a point spread function ($\theta_{\rm psf}$) become dimmer and smaller for the space-based surveys because of the atmosphere scattering for the ground-based survey. The zero-point magnitude ($m_{\rm zp}$) and exposure time ($t_{\rm exp}$) are used to count photons as a signal. $m_{\rm sky}$ and $m_{\rm zp}$ are adjusted to the Johnson-Cousins photometric system.
%The received signal for each telescope is finally optimised applying the transmission efficiency ($\eta_T$).

\section{Microlensing simulation}	\label{sec:micro}

The parallax simulation was based on our previous simulation of FFP microlensing described in our paper \citep{Ban2016}. In this section, we review the FFP simulation with signal-to-noise consideration described in \citep{Ban2016}. The goal was to calculate the expected number of FFP observation by \euclid{}, \wfa{}, and \lsst{} in contrast to  ongoing ground-based surveys such as \moa{} and \ogle{}. The FFP event rate derived in our previous paper is going to be applied to our parallax simulation in this paper.

\subsection{\bes{} galactic model}	\label{subsec:besancon}

We used the stellar data from the \bes{} model version 1112, which was created by \citet{Robin2004, Robin2012a, Robin2012b}. The model comprises the stellar distributions of four populations: thin disc, thick disc, bulge, and spheroid. Each population is modelled with a star formation history, and initial mass function and kinematics are set according to an age-velocity dispersion relation in the thin disc population. For the bulge population, the kinematics are taken from the dynamical model of \citet{Fux1999}. In the bulge, a triaxial Gaussian bar structure is taken to describe its density law. The reddening effect of the interstellar medium (ISM) is considered using the 3D distribution derived by \citet{Marshall2006}.  The model is the same as we used in our previous simulation, and the details are described in \citet{Ban2016}. Here we only mention the parameters of our target area.

\begin{figure}
\centering
\includegraphics[width=3.5in]{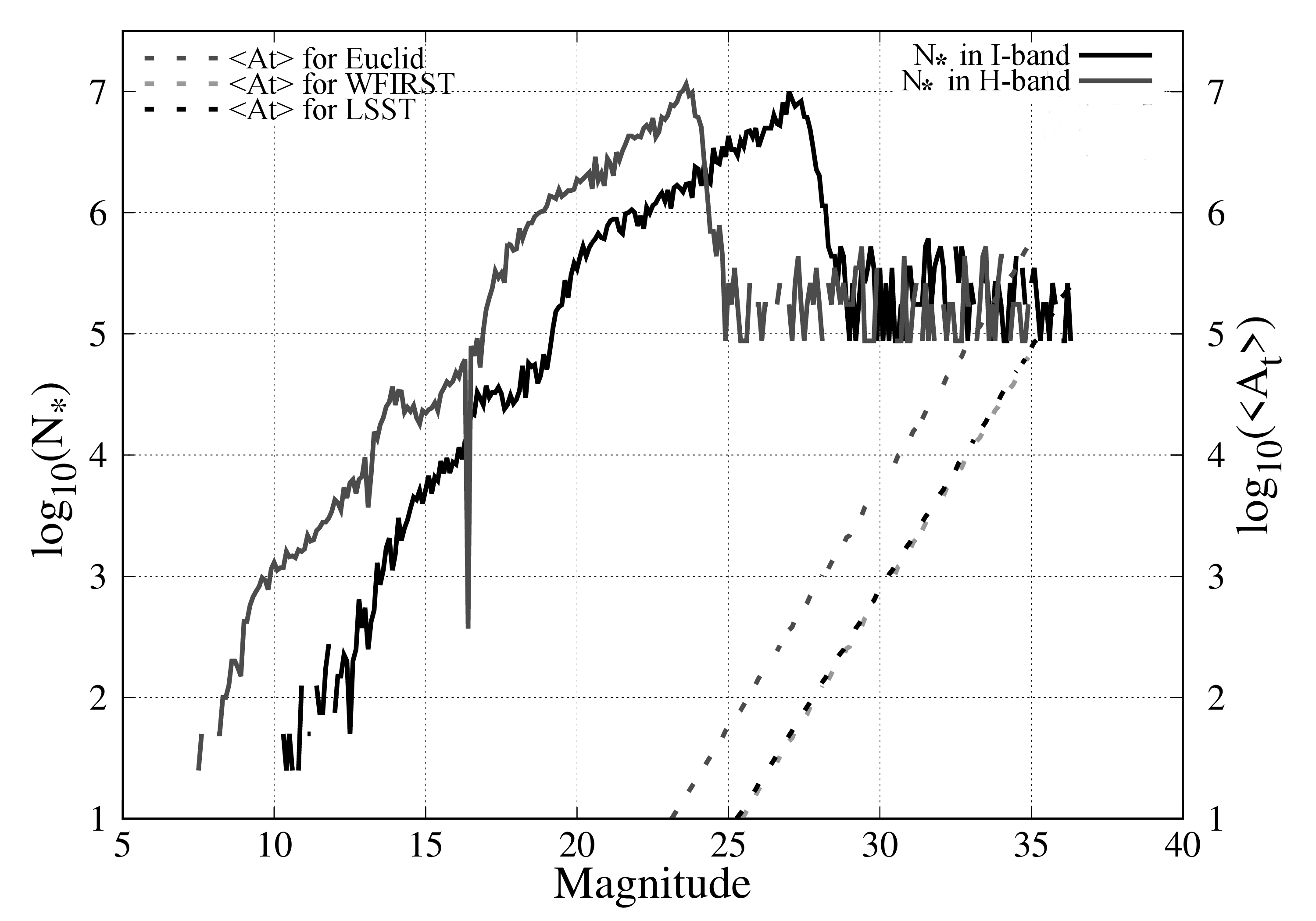} \\
\caption{Graph of stellar population per square degree ($N_*$ labelled at left y-axis) and average threshold impact parameter (<$A_t$> labelled at right y-axis) along with the different magnitude. the magnitude bin is 0.1. The black and dark grey lines are the stellar population in common logarithm for $I$-band and $H$-band, respectively. The black, dark grey and light grey dotted lines are the average threshold amplitude in common logarithm based on the \euclid{}, \wfa{}, and \lsst{} sensitivity, respectively.}
\label{fig:bes-pop}
\end{figure}

\citet{Penny2013} found the discrepancy of the microlensing optical depth value toward the Galactic bulge between the \bes{} model (ver.1106) and observed data in $I$-band and applied a correction factor of 1.8 to their results. In our parallax simulation, we simulate the same observational target $(l, b)=(1^{\circ}, -1.^{\circ}75)$ as \citeauthor{Penny2013} and also apply the correction factor 1.8. This survey target is very close to the ``hot-spot'' of microlensing observation by a ground-based survey (\citealt{Sumi2013}). Table \ref{tab:catals} shows the parameters of the \bes{} model for our research. The catalogues offer a population of about 15.6 million stars within 0.25 $\times$0.25 deg$^{2}$ region centring at our survey target. To gain a statistically reasonable number of stellar data, we divided the stellar catalogues into 4 depending on the magnitude. Since the luminosity function of stars increases significantly towards fainter magnitudes, by invoking a larger solid angle for the simulated brighter stars, we can ensure that they are sampled in the simulated data set, without creating a computationally unfeasible number of fainter stars. Figure \ref{fig:bes-pop} visualises the stellar luminosity functions throughout four catalogues along with the source magnitude bin of 0.1. The average threshold amplitudes based on the \euclid{} and \wfa{} sensitivity for $H$-band and the \lsst{} sensitivity for $I$-band are plotted together. The \bes{} data is successfully providing a smooth population throughout four catalogues. Our criteria of event detectability seem to require that catalogue D stars are strongly amplified to be detected by the telescopes. The details of the detectability test are described in the next section.

\begin{table}
% \centering
\hspace{0.3in}
 \begin{minipage}{140mm}
  \caption{\bes{} catalogue parameters adopted for this work.}
\label{tab:catals}
  \begin{tabular}{@{}lp{1.5in}p{1.0in}@{}}
  \hline
  Main band & $K$-band \\ 
  colour bands & $I-K$, $J-K$, $H-K$ \\ \hline
  survey target & $(l, b)=(1^{\circ}, -1.^{\circ}75)$ \\
  survey region & $0.25\times0.25$ deg$^2$ \\		
  distance range [kpc] & 0-15\\ \hline
  magnitude range & ctlg.A : K = 0-12 \\ 
   & ctlg.B : \hspace*{5.5mm}12-16 \\
   & ctlg.C : \hspace*{5.5mm}16-20 \\
   & ctlg.D : \hspace*{5.5mm}20-99 \\ \hline
  solid angle [deg$^2$] & ctlg.A : 0.0625 \\
   & ctlg.B : $6.8\times 10^{-3}$ \\
   & ctlg.C : $2.1\times 10^{-4}$ \\
   & ctlg.D : $3.6\times 10^{-5}$ \\ \hline
\end{tabular}
\end{minipage}
\end{table}

\subsection{FFP event detectability}	\label{subsec:ffp}

Objects from the four catalogues were used for the source and lens properties, and all combinations of a source and lens were tested. We assumed three FFP mass cases: Jupiter-mass, Neptune-mass, and Earth-mass. Hence, the lens mass was fixed and replaced by these three whilst the distance and proper motion values were taken from the catalogues.

For every source-lens pair, the signal-to-noise ratio was calculated. We defined the detected signal as the number of photons received by a telescope during the exposure time. The noise came from the flux of background unresolved stars, nearby star blending, sky brightness and the photon shot noise from the event itself. Thus, the equation of the signal-to-noise ratio becomes, 
\begin{equation}	\label{eq:sn}
%S/N(t) = \frac{10^{0.2 m_{\rm zp}} \, t^{1/2}_{\rm exp} \, A(t) \, 10^{-0.4m_*}}{\sqrt{10^{-0.4 m_{\rm stars}} + \Omega_{\rm psf}10^{-0.4m_{\rm sky}} + A(t) \, 10^{-0.4m_*}}},
S/N(t) = \frac{10^{0.2 m_{\rm zp}} \, t^{1/2}_{\rm exp} \, A(t) \, 10^{-0.4m_*}}{\sqrt{10^{-0.4 m_{\rm stars}} + 10^{-0.4 m_{\rm blend}} + \Omega_{\rm psf}10^{-0.4m_{\rm sky}} + A(t) \, 10^{-0.4m_*}}},
\end{equation}
where $m_*$ is the apparent magnitude of the source star, $A(t)$ is the microlensing amplitude factor at time $t$, and $m_{\rm zp}$, $t_{\rm exp}$ and $m_{\rm sky}$ are sensitivity-dependent parameters for every observer shown in Table \ref{tab:survs}. $\Omega_{\rm psf}$=$\pi$$\theta_{\rm psf}^2/4$ is the solid angle of the survey point spread function (PSF) where $\theta_{\rm psf}$ is also in Table \ref{tab:survs}. $m_{\rm stars}$ is the combined magnitude contribution of all unresolved sources within the survey target angle (0.25$\times$0.25 deg$^2$), and $m_{\rm blend}$ is the combined magnitude contribution of nearby bright stars around a given target. To determine $m_{\rm stars}$, we have to find the boundary between the resolved and unresolved regimes of our catalogue. Suppose if the baseline magnitude of a given star ($j$) attributes to the flux ($F_j$) and fainter stars than the given star are unresolved, the background noise ($\sqrt{B_{res}}$) can be calculated as the combined flux of those unresolved stars. Subsequently, another signal-to-noise equation is
\begin{equation}	\label{eq:resolvedstar}
\frac{F_j}{\sqrt{B_{res}}} = \frac{10^{0.2 m_{\rm zp}} \, t^{1/2}_{\rm exp} \, 10^{-0.4m_j}}{\sqrt{\ts\Omega_{\rm psf}\sum\limits_{m_i > m_j}\frac{10^{-0.4m_i}}{\ts \Omega_{\rm cat,i}}}},
\end{equation}
where $\Omega_{\rm cat}$ is the solid angle of the \bes{} data catalogue, and the depth of summation ($\sum_{m_i > m_j}$) is dependent on the given star ($j$). We defined the resolved stars satisfy $F_j\sqrt{B_{res}}>3$ for the PSF noise contribution from the unresolved stars. Sorting the stars by magnitude throughout the catalogues, we find the boundary star ($j_{lim}$) which is the faintest resolved star satisfying $F_j\sqrt{B_{res}}>3$. The brighter sources easily satisfy the condition even though the number of fainter stars counted into the noise increases. Once the boundary star ($j_{lim}$) is found, the background noise of the unresolved stars is converted to $m_{\rm stars}$;
\begin{equation}	\label{eq:mstars}
B_{res,lim} = \ts\Omega_{\rm psf} \sum\limits_{m_i > m_{j_{lim}}}\frac{10^{-0.4m_i}}{\ts \Omega_{\rm cat,i}} = 10^{-0.4m_{\rm stars}}.
\end{equation}
Thus, the $m_{\rm stars}$ value is attributed to the \bes{} data distribution and is found for every telescope sensitivity. $m_{\rm blend}$ is also found from the combined flux of stars, but this time, the stars which are brighter than the given target and within PSF range are summed up. 
\begin{equation}	\label{eq:mblend}
B_{blend,j} = \ts\Omega_{\rm psf} \sum\limits_{nearby}\frac{10^{-0.4m_i}}{\ts \Omega_{\rm cat,i}} = 10^{-0.4m_{\rm blend}},
\end{equation}
where the summation limit of ``nearby'' is defined as the brighter stars within the PSF of the target star. Thus, all resolved stars ($m_i <= m_{j_{lim}}$) within the PSF area of the given star are counted. All source stars in the catalogues have the individual value of $m_{\rm blend}$ for every telescope sensitivity. Once we initialise $m_{\rm stars}$ and list $m_{\rm blend}$, the round-robin pairing of the source and lens properties from the \bes{} data is carried out to simulate the detectable events.

For each event, to determine the event detectability, we assume the event must show $S/N$>50 at a peak, and this limit offers the threshold amplitude ($A_t\geq A_{min}$) of the event; hence, threshold impact parameter ($u_{\rm t}$). We can ignore the blending influence due to the lens itself since we assume FFP lenses. 

Once the event is confirmed to be detectable by satisfying above conditions ($S/N>50$), the angular Einstein radius $\theta_{\rm E}$ and event timescale for the given source ($j$) and lens ($i$) pair is given by
  \begin{equation}     \label{eq:theta}
    \theta_{{\rm E},ij} = \sqrt{\frac{4G M_i (D_j - D_i)}{c^2 D_j D_i}},
  \end{equation}
  \begin{equation}      \label{eq:time_ij}
    t_{ij} = \frac{\ts u_{\rm max}\theta_{{\rm E},ij}}{\mu_{ij}},
  \end{equation} 
where $M_i$ is the lens mass and $G$ and $c$ are the gravitational constant and the speed of light. $D_{\rm i}$ and $D_{\rm j}$ represent the lens and source distances, respectively, and $\mu_{ij}$ is the relative lens-source proper motion. 
To find the mean timescale of all detectable event, we define an event occurrence weight ($W_{ij}$), which is a factor of physical lens size and lens speed.
  \begin{equation}     \label{eq:wrate}
  W_{ij} = u_{{\rm t},ij} D_i^2 \mu_{ij} \theta_{{\rm E},ij}.
  \end{equation}
Since we tested all possible source-lens pairs overall four catalogues, the population difference defined by the solid angle  ($\Omega_{cat}$) should also be considered:
\begin{equation}	\label{eq:pop}
p_{ij} = \sum_s \frac{\ts 1}{\ts \Omega_{{\rm cat},s}} \sum_j \sum_l     \sum_{i,D_i < D_j} \frac{\ts P_{FFP,l}}{\ts \Omega_{{\rm cat},l}}.
\end{equation}
$P_{FFP,l}$ is a population ratio of FFPs per star for each catalogue. We assume $P_{FFP}$=1 for all FFP mass cases (i.e. one Jupiter-mass, one Neptune-mass, and one Earth-mass planet per stars), and for all catalogues. This is discussed further in \S\ref{subsec:initial}. These equations perform overall lenses ($i$) drawn from catalogue ($l$) and all sources ($j$) drawn from catalogue ($s$). Thus, the mean timescale ($\langle t \rangle$) is
 \begin{equation}      \label{eq:time}
\langle t \rangle = \frac{p_{ij}W_{ij}t_{ij}}{p_{ij}W_{ij}}.
%\langle t \rangle = \left( \sum_s \frac{\ts 1}{\ts \Omega_{{\rm cat},s}} \sum_j \sum_l     \sum_{i,D_i < D_j} \frac{W_{ij}}{\ts \Omega_{{\rm cat},l}} \right)^{-1} \sum_s \frac{\ts 1}{\ts \Omega_{{\rm cat},s}} \sum_j  \sum_l \sum_{i,D_i < D_j} \frac{\ts W_{ij}}{\ts \Omega_{{\rm  cat},l}} t_{\rm E,ij}.
  \end{equation} 
Note that $\langle t \rangle$ becomes the mean ``Einstein'' timescale $\langle t_E \rangle$ when $u_{\rm max}=1$.

The optical depth for a given source ($\tau_j$) is defined as compiling possible lenses between the source and an observer. The population difference of four catalogues is also considered here. Therefore, the optical depth for a given source ($j$) is
\begin{equation}    \label{eq:tau_j}
   \tau_j=\sum_l \sum_{i,D_i < D_j}  \ts \pi  \theta_{{\rm E},ij}^2\frac{P_{FFP,l}}{\ts \Omega_{{\rm  cat},l}},
\end{equation}
where the equation performs overall lenses ($i$) drawn from catalogue ($l$). $D$ and $\Omega_{\rm cat}$ are the distance and solid angle from the catalogue, respectively. The final optical depth ($\tau$) is the mean value over all possible sources weighted by the ``effectivity'' of the microlensing. Here we define the effectivity as the impact parameter factor ($U_j^{(N)}$) of a given source ($j$) where $N$ varies by the proportionality of the target parameter to the Einstein radius:
 \begin{equation}    \label{eq:uwgt}
   U_j^{(N)} = \frac{\ts \sum_l P_{FFP,l}\Omega_{{\rm cat},l}^{-1} \sum_{i,D_i<D_j} \mbox{min}[1,(u_{{\rm t},ij}/u_{\rm max})^N]}{\ts \sum_l P_{FFP,l}\Omega_{{\rm cat},l}^{-1} \sum_{i,D_i<D_j} 1}.
   \end{equation}
Since $\tau \propto \theta_E^2$, $N=2$ is applied and the final optical depth for all possible sources is expressed as
 \begin{equation}     \label{eq:tau}
    \tau = \left( \sum_s \frac{\ts 1}{\ts \Omega_{{\rm cat},s}} \sum_j U_j^{(2)}\right)^{-1} u_{\rm max}^2 \sum_s \frac{\ts 1}{\ts \Omega_{{\rm cat},s}} \sum_j U_j^{(2)} \tau_j,
    \end{equation}
where the equation is calculated over all sources ($j$) drawn from catalogue ($s$), and we have already assumed $u_{\rm max}$=3 for our telescopes.

 Finally, the source-averaged event rate is given by the optical depth and mean timescale calculated above. The standard formula of the event rate ($\Gamma$) is
 \begin{equation}    \label{eq:eventrate}
   \Gamma = \left[\frac{2}{\pi} \frac{\tau}{\langle t \rangle}\right].
\end{equation} 
So far, we have factorised the timescale and optical depth by the maximum impact parameter to reflect the telescope sensitivity into the event rate. However, there is another way to do this; Eq. \ref{eq:eventrate} can be rewritten as,
 \begin{equation}    \label{eq:eventrate2}
   \Gamma = u_{\rm max}\left[\frac{2}{\pi} \frac{\tau_1}{\langle t_E \rangle}\right],
\end{equation} 
where $\tau_1$ is the optical depth for the $u_{\rm max}$=1 case and $\langle t_E \rangle$ is the mean Einstein timescale. Thus, Eq.(\ref{eq:eventrate2}) is that the event rate for the $u_{\rm max}=1$ case factored by any maximum impact parameter $u_{\rm max}$. 
Finally, the actual number of events per year ($\tilde{\Gamma}$) is given by $\Gamma$$\times$$N_*$ where $N_*$ is the number of source stars for the observation period counted as
   \begin{equation}   \label{eq:sourcenum}
     N_* = \sum_s \sum_{j,U_j^{(1)} > 0} 1.
   \end{equation}
$U_j{(1)}$ is given by Eq.(\ref{eq:uwgt}) with $N=1$.

\subsection{FFP event rate as applied to the parallax simulation}	\label{subsec:ffp-to-prlx}
In \citet{Ban2016}, we simulated a 200 $deg^2$ survey field and mapped the results. The maximum predicted microlensing event rate was recovered in a ``hot-spot'' (Table\ref{tab:ffp-ev}), which defines our simulated field. The ground-based (\lsst{}) sensitivity expects a higher noise level originating from the larger PSF and the unresolved background stars, leading to a lower event rate. We take in to account the rate-weight of the events and the uncertainty of the parameters when computing every formula above. The farther the source is, the more lenses pass the front so that the uncertainty of the optical depth per source gets large. The calculated uncertainty of the event rate is less than 0.01\% for every band and FFP mass so that it is omitted from Table \ref{tab:ffp-ev}. However, this uncertainty does not include the complexities of modelling real-world events, which may increase the sensitivity above what is modelled in unpredictable ways.

In real observation, however, the interference of any other objects and events cannot be treated so simply, hence our method is just ignoring such unexpected interferences and only assuming the ubiquitous causes of uncertainties. Moreover, we assumed S/N>50 at the peak, and the event may be too short to have sufficient indicidual exposures above the threshold signal-to-noise at which an event can be identified. Therefore, as the next step of the microlensing event simulation, we create the time-dependent observation model and simulate parallax observations. The time-dependent model is expected to yield the probability of simultaneous parallax observation that provides the parallax event rate per year multiplied by the annual number of detectable FFP microlensing events.

In the parallax simulation, the source and lens properties are randomly selected unlike a round-robin pairing done in \citet{Ban2016}. The given pair is thrown into the signal-to-noise detectability test ($S/N>$50, Eq.\ref{eq:sn}). If it passes the test, the time-dependent observation model is applied to the event as the simultaneous parallax observation is tried by the expected telescope combinations. The process is repeated until 30,000 events are detected in parallax for every fixed FFP mass (i.e. Jupiter-mass, Neptune-mass, and Earth-mass) to offer a statistically plausible probability of simultaneous parallax. In the next section, we describe the detail of the parallax simulation process and some calculations for further analyses of successful parallax observations by \euclid{} and either \lsst{} or \wfa{}.

\begin{table}
\centering
\hspace{0.3in}
 \caption{FFP event rate ($\tilde{\Gamma}_{FFP}$ [events year$^{-1}$ deg$^{-2}$]) at $(l, b)=(1^{\circ}, -1.^{\circ}75)$ retrieved from the data used in \citet{Ban2016}. Note that the event rate is under the solo-observation of the telescope without considering the operation seasons and time. For \lsst{}, the night-time lasting (7.5h) and fine-weather ratio (65.89\%) was applied.}
\label{tab:ffp-ev}
  \begin{tabular}{llll}%{@{}lp{1.5in}p{0.5in}p{0.5in}@{}}
  Lens mass & \euclid{} & \wfa{} & \lsst{} \\ \hline
  Jupiter & 2045 & 2026 & 377 \\
  Neptune & 475 & 470 & 87 \\
  Earth & 114 & 114 & 21 \\
\end{tabular}
\end{table}
%  Lens mass & $I$-band & $H$-band \\ \hline
%  Jupiter & 1855 & 2070 \\
%  Neptune & 427 & 480 \\
%  Earth & 98 & 111 \\
%  Jupiter & 2120 & 2070 \\
%  Neptune & 490 & 480 \\
%  Earth & 112 & 111 \\

\section{Simultaneous parallax observation}	\label{sec:prlx}
In this section, we describe the structure of the parallax simulation. The time-dependent model offers the light curves for every simulated event, and we use the light curves to determine the parallax detectability. The goal is to derive the parallax event rate and the accuracy of the lens-mass estimation from the differential light curves.

\begin{figure*}
{\centering
\begin{tabular}{cc}
Object positions in 2D & Sample light curve \\
\hspace{-0.2in}\includegraphics[width=4in]{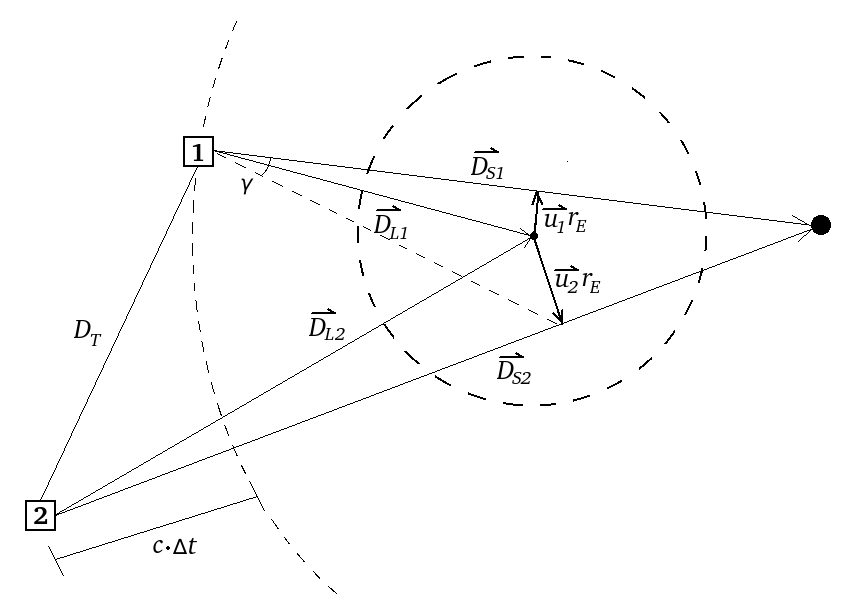}  &
\hspace{-0.1in}\includegraphics[width=3in]{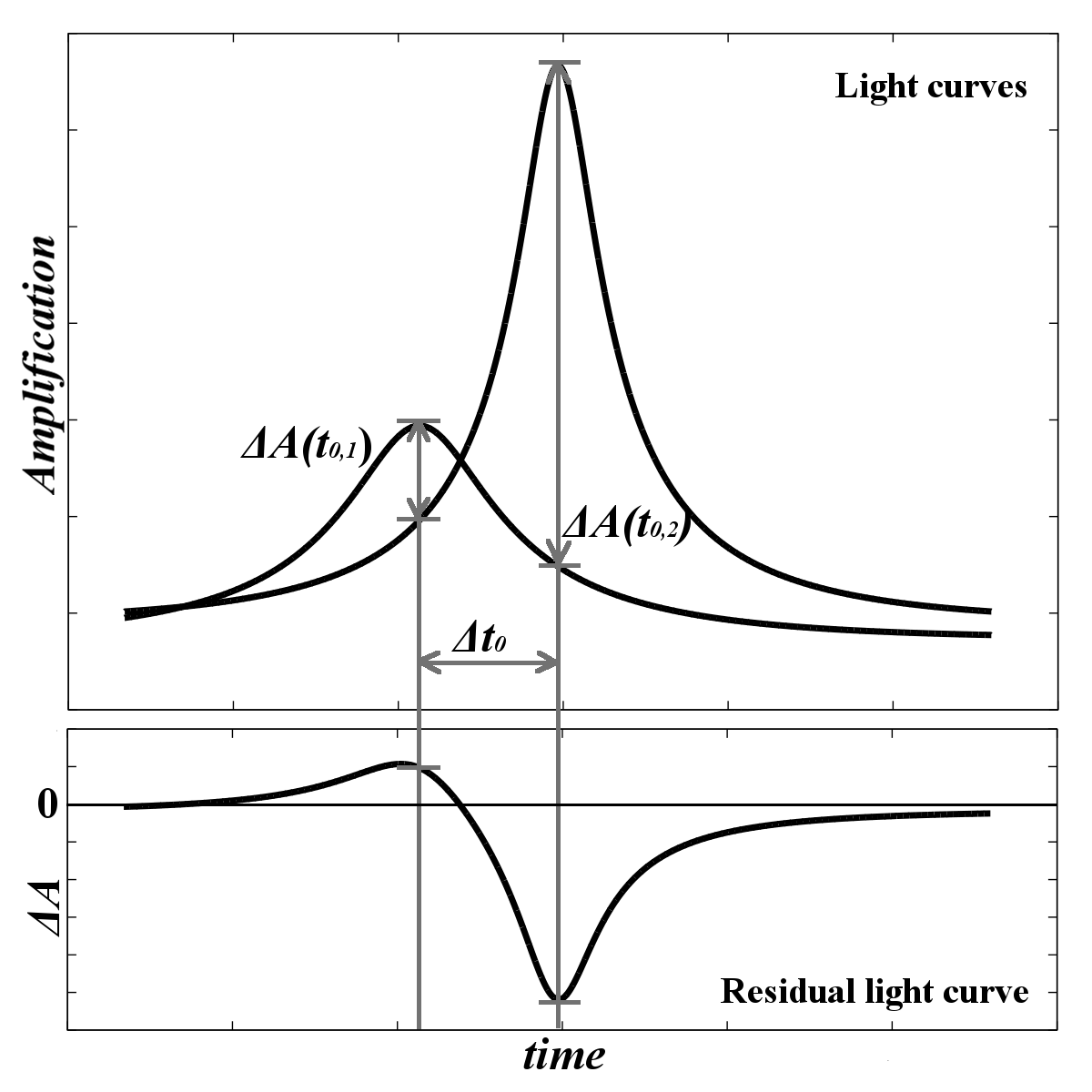} \\
\end{tabular}
}
\caption{Schematic images of simultaneous parallax observation concept. {\it Left panel}: A two-dimensional geometrical arrangement for a source, lens and two observers. The large black dot is the source star, the small black dot is the lens object, and the numbers in boxes represent two observers. The dashed circle centring at the lens shows the Einstein ring. $\vec{D_S}$ and $\vec{D_L}$ are the line-of-sight to the source and lens from each observer, respectively. $\vec{u}$ is vectorial impact parameter and $r_E$ represents the Einstein radius in length. $D_T$ is the observer separation. The dashed curve represents the light horizon of the event when it is detected by Observer 1; an extra time ($\Delta t$) is necessary for Observer 2 to detect the same radiation which is detected by Observer 1 at this moment. The dashed line from Observer 1 to $\vec{D_{S2}}$ is just a supportive line to visualise the parallactic angle $\gamma$ that also varies in a function of time. {\it Right panel}: Sample light curves of simultaneous parallax observation in units of amplitude factor (top) and differential amplitude between two observers (bottom). We define $t_{0,1}$ and $t_{0,2}$ as the light curve peak of each observer and so $\Delta t_0$ as the difference in time of maximum amplification. We will use the absolute value of $\Delta A$ in later calculations. }
\label{fig:prlx-geo}
\end{figure*}

\subsection{Configuration of an event simulation}	\label{subsec:initial}
We applied some random values for the parameters in our simulation: source-lens pair selection, zero-time ($t_{event}=0$) reference and minimum impact parameter. The randomness was controlled by the proper probability if the distribution is not uniform.

\subsubsection{Source-lens pair selection} \label{subsubsec:random1}
First, a source and lens are randomly chosen from four catalogues. The probability of data selection is controlled based on the stellar population defined by solid angles. The stellar population ratio is 1:14:455:3200 from catalogue A to D (brighter to fainter). Hence, more faint sources will be chosen. A lens is selected in the same way, but it must be at a closer distance than the source. As in \cite{Ban2016}, we assume three fixed planetary lens cases: Jupiter-mass, Neptune-mass, and Earth-mass, so that the lens mass is replaced with them and only the distance and proper motion were referred from the catalogue.

The FFP population is assumed to be 1 FFP per star. \cite{Sumi2011} first estimated that the Jupiter-mass FFP population is about twice as large as the main-sequence (MS) star population. Later, \cite{Mroz2017} recalculated the population, resulting in an upper limit at 95\% confidence of $\sim$0.25 FFPs per MS star. The population of FFPs of any given mass is still effectively unknown.  Hence, for simplicity, we adopt an FFP frequency of one FFP of each mass (Jupiter-mass, Neptune-mass, and Earth-mass) per star, with the true number of expected planets to be normalised by the end-user once these are ascertained.

\subsubsection{Zero-time reference} \label{subsubsec:random2}
Second, two angles are randomly chosen for observers' location. One is the positioning angle of the Earth on the orbital plane in the heliocentric co-ordinates. The other is the phase position of \euclid{} on the orbit around L2 in the terrestrial sky reference frame (i.e. 2D with the L2 origin). We set these positions at $t_{event}=0$ in our time-dependent observation model when the reference observer on the Earth detects the maximum amplitude of the event. Thus, the ingress ($t_{event}<0$) and egress ($t_{event}>0$) portions of the event are simulated with respect to the $t_{event}=0$ positions. In case of \euclid{}-\wfa{} combination, we assume that \wfa{} in the Halo orbit shares the \euclid{} orbit around L2. The relative motion of the source, lens and telescopes in the 3D space varied in our simulation. We considered it was too much to make the \wfa{} position random in addition to the \euclid{} position. Under the shared orbit, all phase differences are possible to be numerically handled by the distribution probability. Besides, the small phase difference (i.e. small separation between \euclid{} and \wfa{}) cannot take advantage of the parallax observation. We decided to take the 90$^{\circ}$ phase difference in our model. For the \euclid{}-\lsst{} combination, the Earth's rotation is not operated in our simulation. Instead, we assumed that the observable night is lasting 7.5 hours and the weather fineness is $65.89\%$ as it is mentioned in \S\ref{subsec:tele-kine}. Once the positions $t_{event}=0$ are determined, the solar aspect angle (SAA) is examined. As shown in \$\ref{subsec:tele-kine}, the telescope position must satisfy 90$^{\circ}$<SAA<120$^{\circ}$ for \euclid{} and 54$^{\circ}$<SAA<126$^{\circ}$ for \wfa{}. Our time-dependent model allows changing SAA during the event along with the telescope motion. However, we assume that \euclid{} and \wfa{} must be within their SAA range at $t_{event}=0$. The edge-of-SAA issue is not so serious because the SAA shift during the event will be quite small within the FFP event duration and we will also have the simultaneous parallax duration limit mentioned later in \S\ref{subsec:prlx-obs}.

\subsubsection{Minimum Impact parameter} \label{subsubsec:random3}
Third, the impact parameter ($u_{\oplus}(t)$) at $t_{event}=0$ is randomly chosen; where the subscript symbol ($\oplus$) means the reference observer on the Earth. Note that we assumed the reference observer detects maximum amplitude at $t_{event}=0$, and $u_{\oplus}(t_{event}=0)$ becomes the minimum impact parameter for the reference observer. The minimum impact parameter must be less than the threshold impact parameter  ($u_{\oplus}(t_{event}=0)<u_{t,\oplus}$) found in the process of event detectability discussed in $\S$\ref{subsec:ffp}. The distribution probability of $u_{\oplus}(t_{event}=0)$ is proportional to ${r^2 : r \in u_{t,\oplus}}$ where $r$ is the distance from the lens centre in units of Einstein radii. Whist the coordinates of the source refer to the \bes{} data, the coordinates of the lens are calculated after $u_{\oplus}(t_{event}=0)$ is randomly chosen to yield a microlensing event.

\subsubsection{Statistics of the simulation length} \label{subsubsec:random1}
According to our convergence test for a Monte-Carlo method, 30,000 detectable parallax events is enough amount to gain statistically plausible estimate of the parallax event rate through the simulation. Again, we have 9 runs in total: 3 telescope combinations (\euclid{}-\wfa{} with the Halo orbit, \euclid{}-\wfa{} with the geosynchronous orbit, and \euclid{}-\lsst{}) times 3 fixed lens-mass cases (Jupiter-mass, Neptune-mass, and Earth-mass). Each run requires 30,000 detectable parallax events. Accordingly, it was necessary to test 4-11 times more events to obtain 30,000 detectable parallax events, since we set some conditions of detectability as follows.

\subsection{Parallax observation}	\label{subsec:prlx-obs}

Figure \ref{fig:prlx-geo} shows the observational setup for simultaneous parallax observation. The left panel is a schematic 2D image of the event object arrangement, and the right panel shows sample light curves with their difference (residual light curve). The parallactic angle $\gamma$ helps to identify the lens mass and distance from the light curves. The value $\gamma$ is determined as
 \begin{equation} \label{eq:prlx_angle}
    \gamma(t) = \left|\vec{u_1}(t)-\vec{u_2}(t)\right|\theta_E
 \end{equation}
 where $\vec{u_1}(t)$ and $\vec{u_2}(t)$ are impact parameters of every observer in a function of time. In real observations, the observed amplitude factors of each telescope represent these impact parameters. \cite{Gould2013} stated the mass and distance equations with microlensing parallax ($\pi_E$) and their equations are rewritten by our parallactic angle as
 \begin{equation} \label{eq:prlx_mass}
    M_L =\frac{\theta_E^2 D_T}{\kappa \gamma} \hspace{0.2in} {\rm where}  \hspace{0.2in} \kappa\equiv\frac{4G}{c^2AU}\sim8.1\frac{{\rm mas}}{M_{\odot}},
 \end{equation}
  and
 \begin{equation} \label{eq:prlx_signal}
    D_L = \frac{D_S D_T}{\gamma D_S+D_T},
 \end{equation}
where these symbols correspond to those in Figure \ref{fig:prlx-geo}. $D_S$ is a standardised distance of the source from the Sun. Note that the impact parameters and parallactic angle time-dependent quantities that will be computed from the light curves. The observer distance ($D_T$) causes a time gap ($\Delta t$) of the arriving signal. As we described in \S\ref{subsec:initial}, we set the reference observer on the Earth, and the time gap for space-based observers are carefully considered in our time-dependent model.

To analyse the parallax observation, we define a parallax signal ($S$) from the residual light curve as
  \begin{equation} \label{eq:prlx_signal}
    S = \Delta A_{max} \times T,
  \end{equation}
where $\Delta A_{max}$ is the maximum absolute value of the differential amplitude, and $T$ is the duration over which the amplitude seen by both telescopes exceeds $A_t$, in hours. Besides, $T$ should be at least 1 hour for given telescopes' cadences of 10-20 minutes. Note that a ``cadence'' for microlensing surveys is usually used in the meaning of the interval between observations/shuttering. Thus, we can instead state that both telescopes require at least three exposures with amplitude above $A_t$ to identify a simultaneous detection of an event. We also consider the noise level of differential amplitude calculated as
  \begin{equation} \label{eq:prlx_errfunc}
    D = \left(\frac{\Delta A(t)}{\sigma_{\Delta A(t)}}\right)_{max},
  \end{equation}
  \begin{equation} \label{eq:prlx_err}
    \sigma_{A_i(t)} = \sqrt{A_i(t)\times10^{-0.4(m_{*,i}-m_{zp,i})}},
  \end{equation}
where $A_i(t)$ is the observed magnification for each telescope. $m_{*,i}$ and $m_{zp,i}$ are the source magnitude and zero-point magnitude in the corresponding photometric band of each telescope taken from Table \ref{tab:survs}. We assume $\sigma_{A_i(t)}\geq3\times10^{-4}$ for every moment $t$ and require D > 5 for a detectable parallax event. 

Finally, we will find the parallax event rate by
  \begin{equation} \label{eq:prlx_er}
    \tilde{\Gamma}_{parallax} = P_{parallax}\times\tilde{\Gamma}_{FFP},
  \end{equation}
where $\tilde{\Gamma}_{FFP}$ is the event rate only from the FFP simulation described in $\S$\ref{subsec:ffp}, and  $P_{parallax}$ is the rate weighted probability of detectable FFP parallax observation to all tested FFP events.
  \begin{equation} \label{eq:prlx_prob}
    P_{parallax} = \frac{\sum\limits_{ij}W_{ij}}{\sum\limits_{all}W_{all}},
  \end{equation}
where $W_{ij}$ is the rate weight value of detectable parallax (see Eq.\ref{eq:wrate}). Note that the SAA limit will constrain detectable events to within two 30-day periods around the equinoxes of each year. $W_{all}$ is the rate weight of all detectable FFP microlensing event through a year which satisfy the S/N>50 limit. Hence, the summation of $W_{all}$ contains events out of SAA, events observed by either telescope within SAA (i.e. no parallax observation), events observed by both telescopes within SAA but failed our parallax detectability limit, and detectable parallax events. The input catalogues enough contain data to operate Monte-Carlo method by repeated random selection of a source-lens pair followed by the random selection of positioning angles and minimum impact parameter.

\section{Results}	\label{sec:res}
The simulation finally provides some mapped results and the numerical result of FFP parallax observation probability. Note that these results are offered for each lens mass case (Jupiter-mass, Neptune-mass, and Earth-mass) under every combination (\euclid{}-\wfa{} and \euclid{}-\lsst{}). Moreover, both the Halo orbit at L2, which is now the official decision, and geosynchronous orbit for \wfa{} are simulated. In \S\ref{subsec:res_pairs} and \S\ref{subsec:res_chara}, we are going to analyse the tendencies of detectable parallax observation with the fraction of detectable events distribution maps; which indicates the percentage probability of the event binned by the parameters shown in the axes among the detectable parallax events (i.e. total detectable events in our simulation is 30,000 events and the percentage is about it). The fraction of detectable events was taken as the rate-weighted probability (see \S\ref{eq:wrate}) and shown as a map. The mapped results of \euclid{}-\wfa{} combination with the Halo orbit at L2 is going to be shown whilst the others are omitted: this is because we confirmed that the mapped results of \euclid{}-\wfa{} combination with geosynchronous orbit showed similar patterns since their separation is the only different condition. The mapped results of the \euclid{}-\lsst{} combination is shown as a differential distribution map from the \euclid{}-\wfa{} combination with the Halo orbit at L2 to clarify their differences. The numerical result of parallax probability for all combinations and cases are finally described. In \S\ref{subsec:res_er} and \S\ref{subsec:res_mass}, we discuss the event rate of parallax observation and the plausibility of our simulation from the view of mass estimation from the output light curves.

\begin{figure*}
{\centering
\includegraphics[width=6.5in]{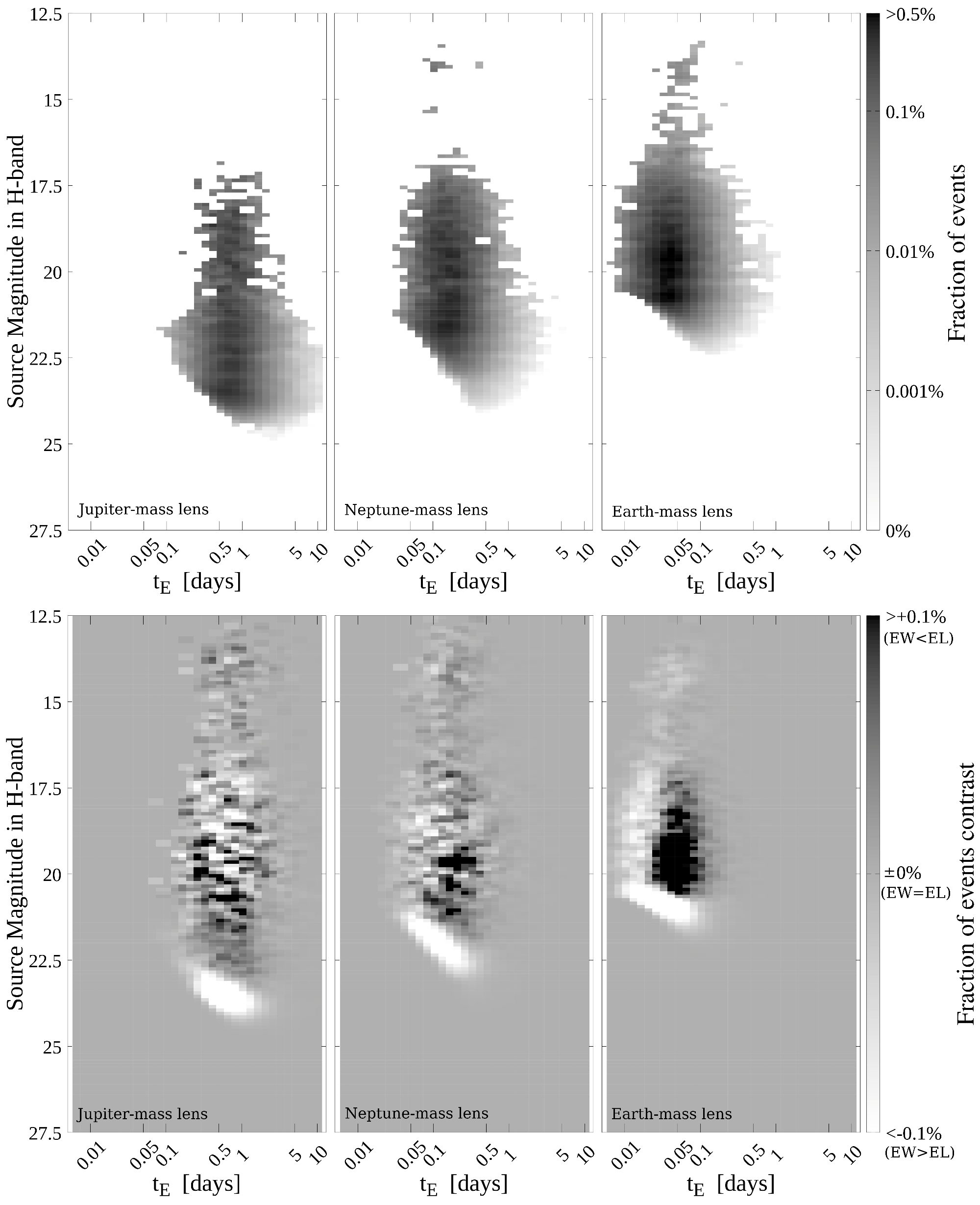} \\
}
\caption{Fraction of detectable events distribution maps for the Einstein timescale and source magnitude in H-band combination with the bin of $\Delta t_E=0.1$ day in logarithmic calibration and $\Delta H=0.1$ per square-degree. {\it Top}: The fraction of detectable events distribution from the  \euclid{}-\wfa{} combination. {\it Bottom}: The difference in the distribution of detected events between the \euclid{}-\wfa{} and \euclid{}-\lsst{} combinations (differential distributions). Note that the original distribution maps are similar to each other on the appearance and the range of fraction. The blacker regime means that relatively more detections will be recovered by the \euclid{}-\lsst{} combination and the whiter regime is more detections by the \euclid{}-\wfa{} combination.}
\label{fig:th}
\end{figure*}

\begin{figure*}%[!htbp]
{\centering
\includegraphics[width=6.5in]{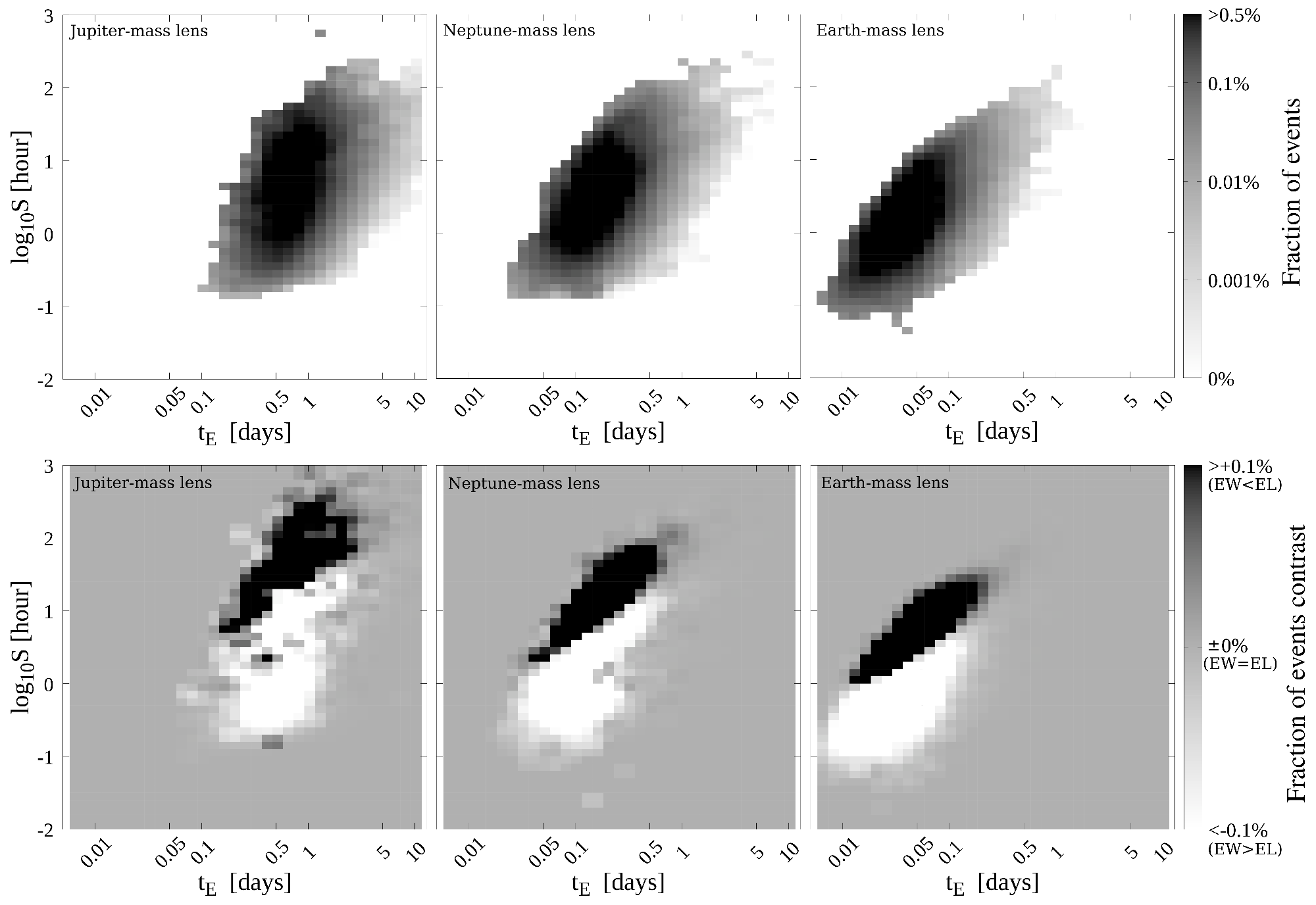} \\
}
\caption{As Figure \ref{fig:th}, but showing the fraction of detectable events distribution maps for the Einstein timescale and parallax signal ($S$) combination with the bin of $\Delta t_E=0.1$ day in logarithmic calibration and $\Delta log_{10}S=0.1$ hour.}
\label{fig:ts}
\end{figure*}

\begin{figure*}%[!htbp]
{\centering
\includegraphics[width=6.5in]{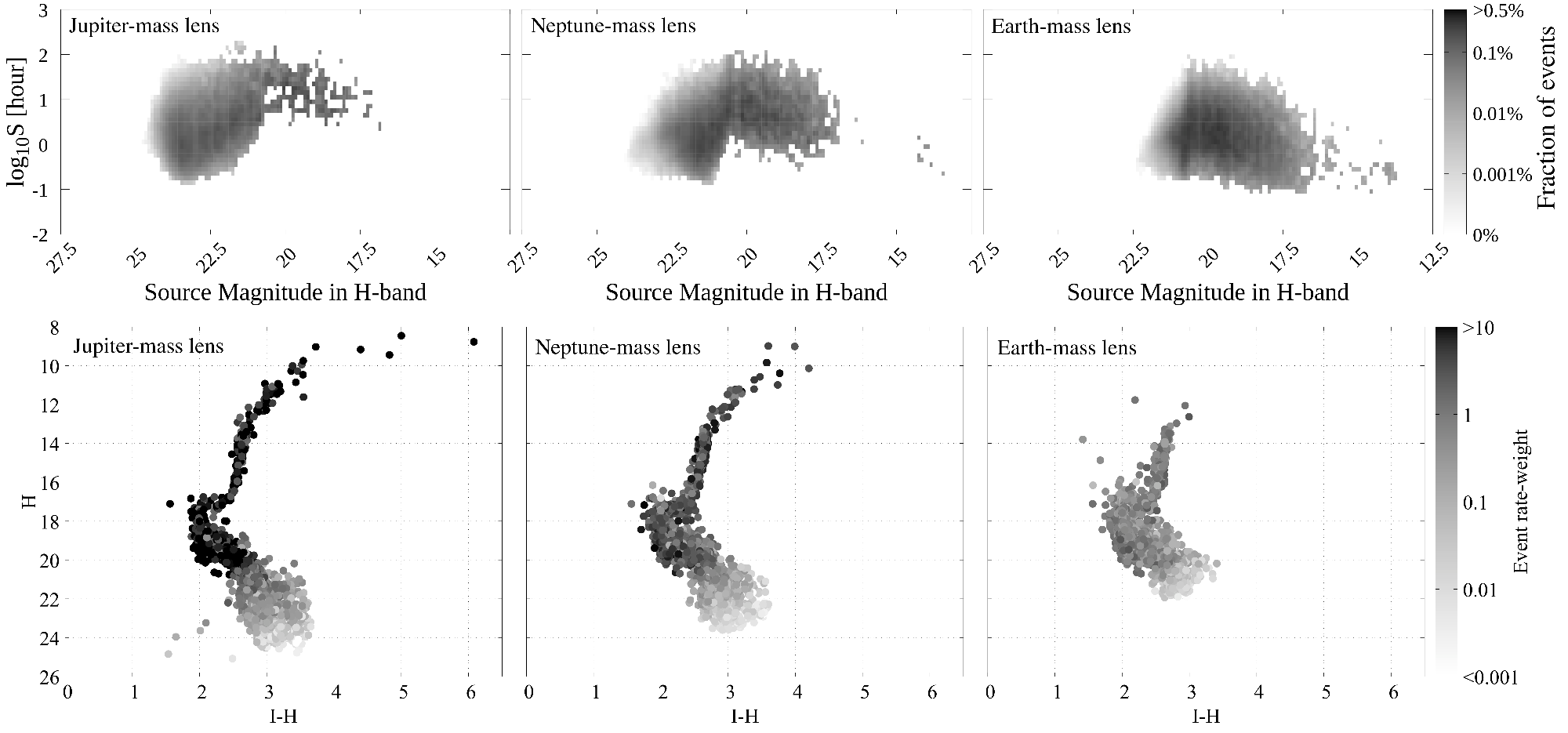} \\
}
\caption{Fraction of detectable events distribution maps for the source magnitude in $H$-band and parallax signal ($S$) combination with the bin of $\Delta H=0.1$ per square-degree and $\Delta log_{10}S=0.1$ hour and the HR diagram of these source stars. {\it Top}: The fraction of detectable events distribution for from the  \euclid{}-\wfa{} combination. {\it Bottom}: The source star origin in HR diagram; where the event rate weight is from Eq.(\ref{eq:wrate}). Note that the \euclid{} and \lsst{} are approximated as observing in $H$-band and $I$-band, respectively.}
\label{fig:hs-hr}
\end{figure*}

\subsection{Distribution of source-lens pairs} \label{subsec:res_pairs}
The population of catalogue stars is almost a continuous distribution in magnitude, but it is true that the faint star population (main-sequence (MS) stars) is much larger than bright stars (i.e. Red Giant Branch (RGB) stars). The faint sources, therefore, dominate the 30,000 simulated parallax events. Figure \ref{fig:th} shows the distribution of the fraction of detectable events in Einstein timescale and source magnitude in $H$-band.  Because both combinations provided similar distribution maps, here we only show the maps of the \euclid{}-\wfa{} combination for the top panels of Figure \ref{fig:th} and the residual fraction between two combinations (i.e. the fraction recovered in the \euclid{}-\wfa{} (EW) combination subtracted from the \euclid{}-\lsst{} (EL) combination) for the bottom panels to see the distribution difference more effectively.

Events associated with each FFP mass concentrate near a particular Einstein timescale. The timescale corresponds to the mean Einstein timescale for every FFP mass with which event is observed solely \citep{Ban2016}. For example, the canonically assumed Jupiter-mass FFP event will have $\theta_E\sim0.03$ mas and $t_E\sim$2 days, and these canonical values are proportional to the square root of the lens mass \citep{Sumi2011}. The mean value from our simulation is shorter than the canonical timescale. The reason is that our simulation likely provided a smaller lens than the canonical size. As Eq.(\ref{eq:theta}) shows, the source and lens distance combination ($(D_j-D_i)/D_jD_i$) determines the size of the Einstein radius for a fixed lens mass. A canonical Jupiter-mass event with $t_E$$\sim$2 days is usually assumed to have $D_j=8$ kpc and $D_i=4$ kpc; which gives $(D_j-D_i)/D_jD_i$=0.125 kpc$^{-1}$. We confirmed that about 97\% of our 30,000 detectable parallax events gave less than 0.125 kpc$^{-1}$. The same tendency appears for the Neptune-mass and Earth-mass cases. Further discussion about the source and lens distance relation will be made in the later paragraph with the lens distance distribution maps.

The lower limit of the source magnitude increases as the FFP mass decreases because a low-mass source requires a small threshold impact parameter to yield a large amplitude and therefore the less-massive lens cannot pass our parallax detection limit of $T>$1 hour. The diagonal cut of the bottom-left edge of the distribution occurs for the same reason, and the effect of short timescale is more strictly appearing. The differential distributions shows that the \euclid{}-\lsst{} combination rises the lower-limit of the source magnitude because the ground-based sensitivity of \lsst{} requires a smaller threshold impact parameter (i.e. higher amplitude) to be observed. Due to the more strict cut-off by the \lsst{} sensitivity and the stellar population, the distribution of the fraction of detectable events concentrates more on $H\sim20$ sources in the \euclid{}-\lsst{} combination than that of the \euclid{}-\wfa{} combination. The differential distributions of Earth-mass lenses specifically show how the limiting timescale differs between two telescope combinations; the \euclid{}-\wfa{} combination can detect shorter-timescale events than the \euclid{}-\lsst{} combination. The influence of the ground-based sensitivity and our parallax detectability limit of $T>$1 hour caused such a clear boundary between two regimes where the distribution is concentrated by the \euclid{}-\wfa{} combination and the \euclid{}-\lsst{} combination. Consequently, the Jupiter-mass lens and Neptune-mass lens cases do not clearly show the gap of the timescale limit between two combinations.

Figure \ref{fig:ts} shows the distribution of the fraction of detectable events in Einstein timescale and parallax signal ($S$ in logarithmic scale, Eq.(\ref{eq:prlx_signal})) combination. The fraction distribution shows that the range of $S$ does not vary depending on the lens mass despite the lens mass determining the Einstein radii, hence Einstein timescale. The lower limit of $S$ can be attributed to our duration limit for the parallax detectability ($T>$1 hour). On the other hand, the upper limit can vary with the Einstein timescale, but we require enough ``distinguishability'' of parallax light curves compared to the noise level (D > 5, see Eq.(\ref{eq:prlx_errfunc})). The larger the Einstein radius and the longer the duration, the more difficult it is to identify differences in the two light curves. Consequently, none of the FFP lens-mass cases exceeds $log_{10}S\sim2.5$. The differential distributions indicates that the \euclid{}-\lsst{} combination is sensitive to larger parallactic angles than the \euclid{}-\wfa{} combination. The higher noise of the ground-based survey requires a larger differential amplitude between the \euclid{} light curve and \lsst{} light curve to satisfy the noise level limit (D > 5) in our simulation. The boundary between the dominant regions of the \euclid{}-\wfa{} combination and the \euclid{}-\lsst{} combination illustrates the relation of $S\propto t_E$.

Figure \ref{fig:hs-hr} shows the relation between $H$-band magnitude and $S$ (top), and the position of source stars on the $I-H$ colour--magnitude diagram (bottom). The fraction distribution shows that $S$ reaches a maximum at a source magnitude of $H\sim$20.5 and decreases to both brighter and fainter source regimes. The peak in source magnitude can be explained using Figure \ref{fig:bes-pop}. Sources with $H>$20.5 mag numerically dominate the source population, but an event requires strong magnification to become detectable. Sources with $H<$20.5 mag are comparatively rare, and finite source effects decreases the detectability of $S$. The near the $H\sim$20.5 mag boundary are common and can easily satisfy the S/N$>$50 criterion without strong amplitude (i.e. without a small impact parameter).

\begin{figure}
{\centering
\vspace{-0.15in}\includegraphics[width=5.4in]{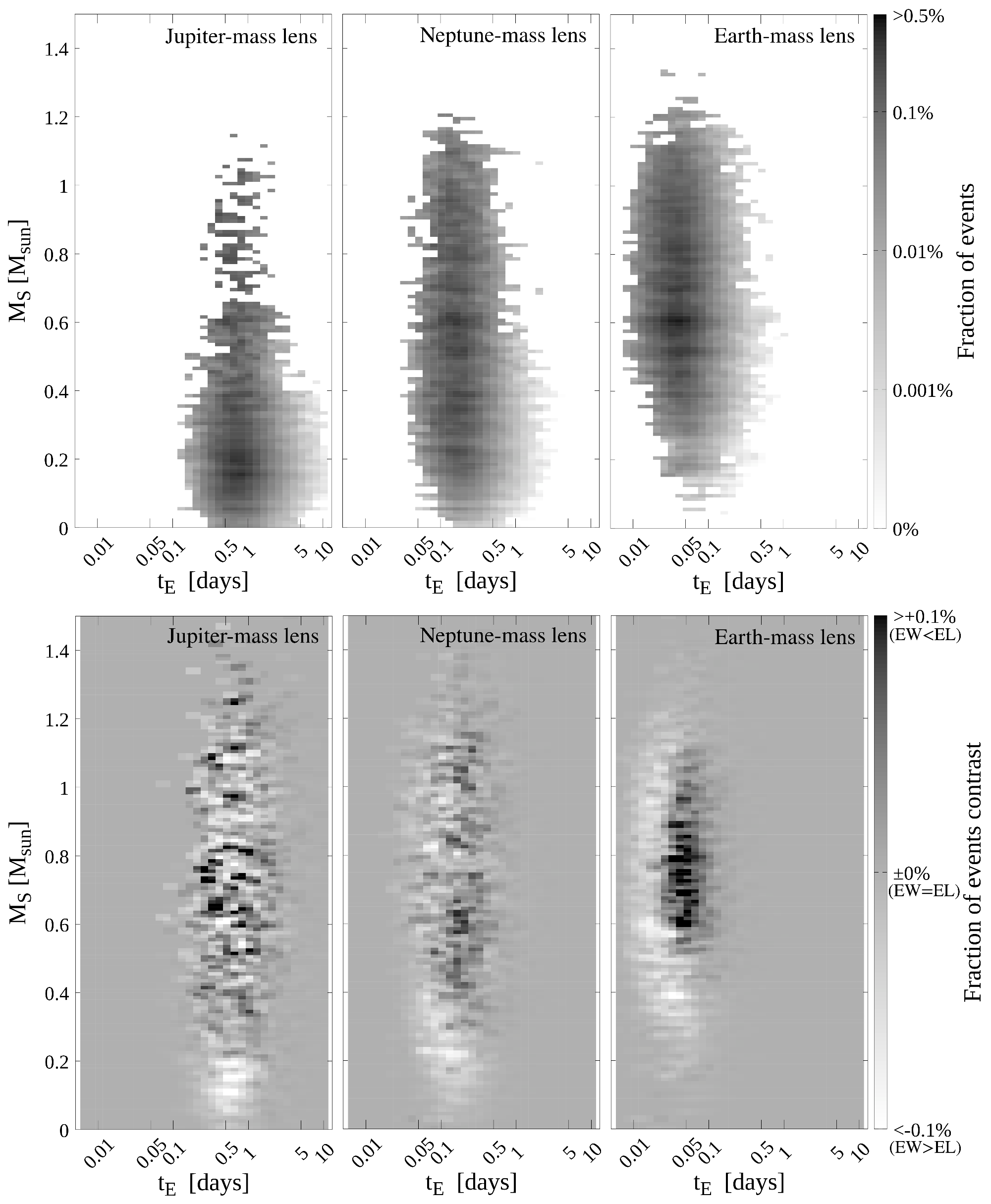} \\
}
\caption{As Figure \ref{fig:th}, but showing the fraction of detectable events distribution maps for the Einstein timescale and source mass combination with the bin of $\Delta t_E=0.1$ day in logarithmic calibration and $\Delta M_S=0.1$ solar-mass.}
\vspace{5in}
\label{fig:tm}
\end{figure}

\begin{figure}
\centering
\vspace{6.5in}
\includegraphics[width=3.4in]{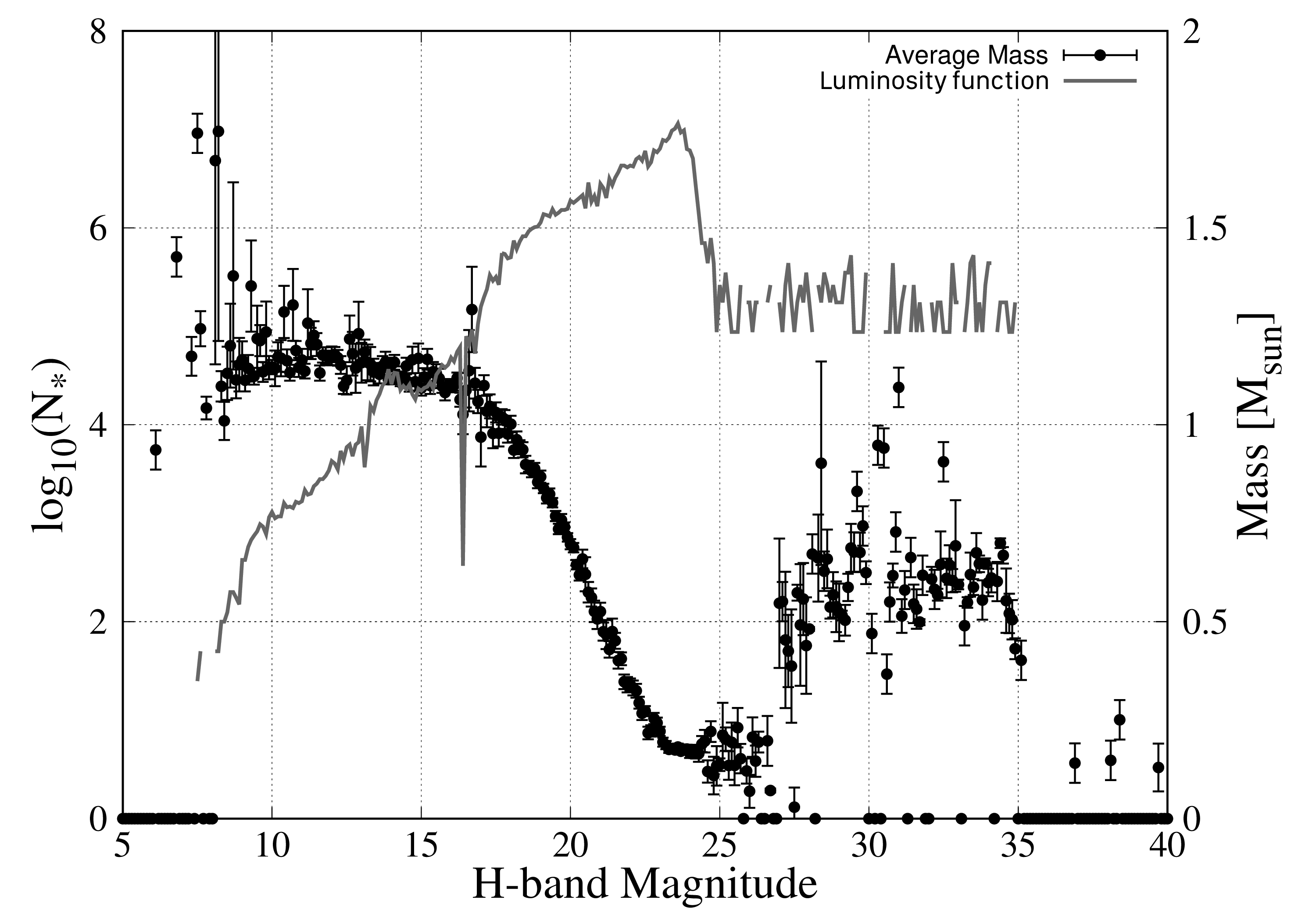}
\vspace{-0.2in}
\caption{The $H$-band luminosity function (grey line, left y-axis) and average mass (black points, right y-axis), as computed by the \bes{} model, in 0.1 mag bins. The error bars represent the maximum-minimum range of stellar masses within each magnitude bin.}
\label{fig:Nmass}
\end{figure}

Figure \ref{fig:tm} shows the distribution of the fraction of detectable events in Einstein timescale and source mass combination. Figure \ref{fig:Nmass} is a supportive plot, showing the stellar mass and the luminosity function from the whole \bes{} catalogue. Detectable events involving Jupiter-mass FFPs typically involve less-massive source stars than events involving Earth-mass FFPs. Since the MS stars have the power-law relationship between luminosity and mass, low mass sources are faint sources which require significant amplitude to satisfy the signal-to-noise detectability limit (S/N>50). The Earth-mass lens events for these faint sources are likely cut-off by our parallax duration limit ($T>$1 hour) due to the small lens size. Hence, the fraction of detectable events involving massive source stars increases for Earth-mass FFPs whilst the fraction for Jupiter-mass FFPs reflects the source population more directly. According to Figure \ref{fig:Nmass}, the majority of faint stars from catalogue C and D were $\leq1M_{sun}$ whose population was much larger than the brighter stars. The differential distributions shows almost the same tendencies as Figure \ref{fig:th}; the \euclid{}-\wfa{} combination can reliably detected parallax-induced differences in the light curves from lower-mass lenses than the \euclid{}-\lsst{} combination, and the Earth-mass lens case shows the timescale limit of the \euclid{}-\lsst{} combination due to the ground-based sensitivity and the parallax detectability limit of $T>$1 hour.

\begin{figure}
{\centering
\vspace{-0.15in}\includegraphics[width=5.4in]{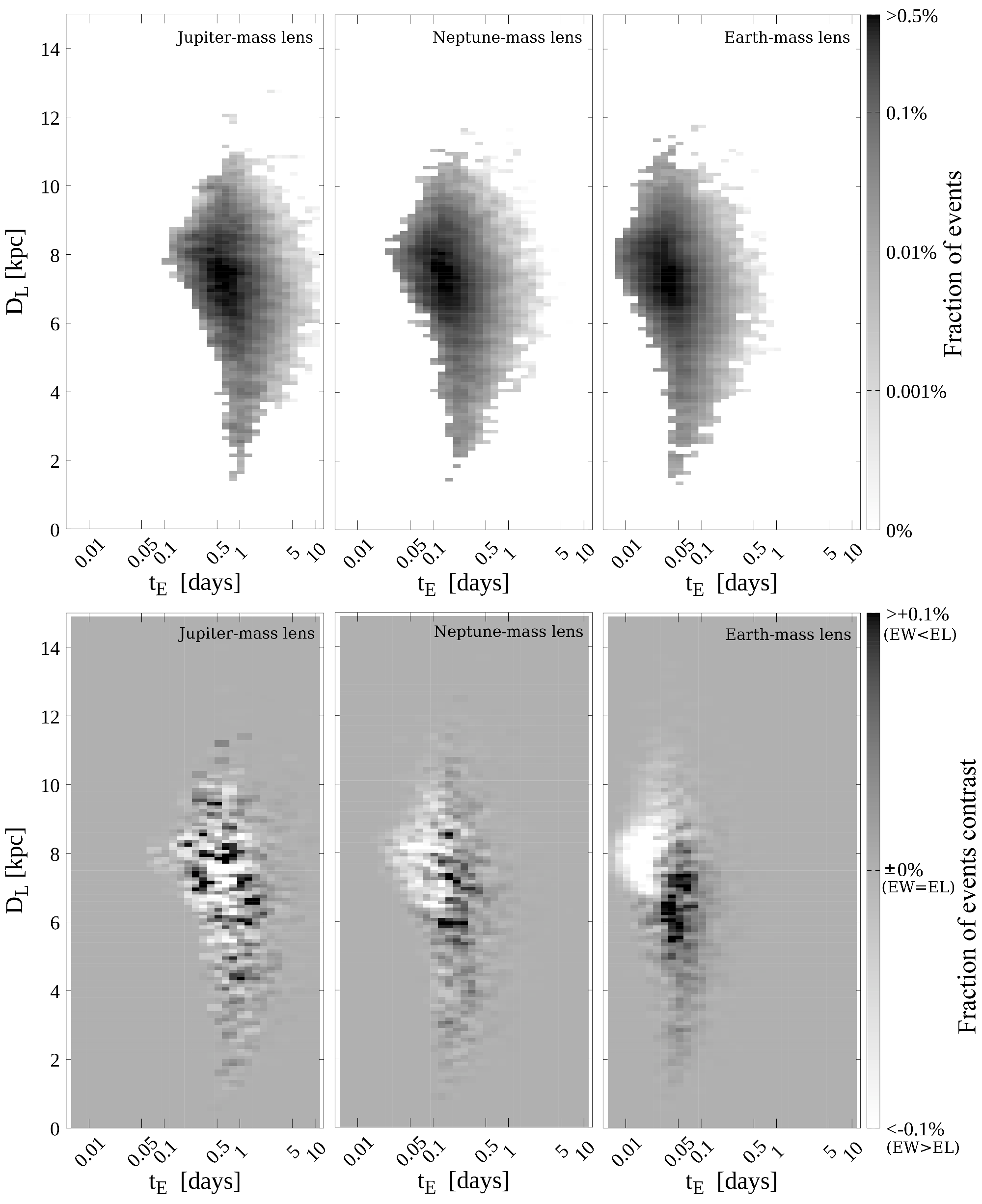} \\
}
\caption{As Figure \ref{fig:th}, but showing the fraction of detectable events distribution maps for the Einstein timescale and lens distance combination with the bin of $\Delta t_E=0.1$ day in logarithmic calibration and $\Delta D_L=0.1$ kpc.}
\vspace{5in}
\label{fig:td}
\end{figure}

\begin{figure}
\centering
\vspace{6.5in}
\includegraphics[width=3.4in]{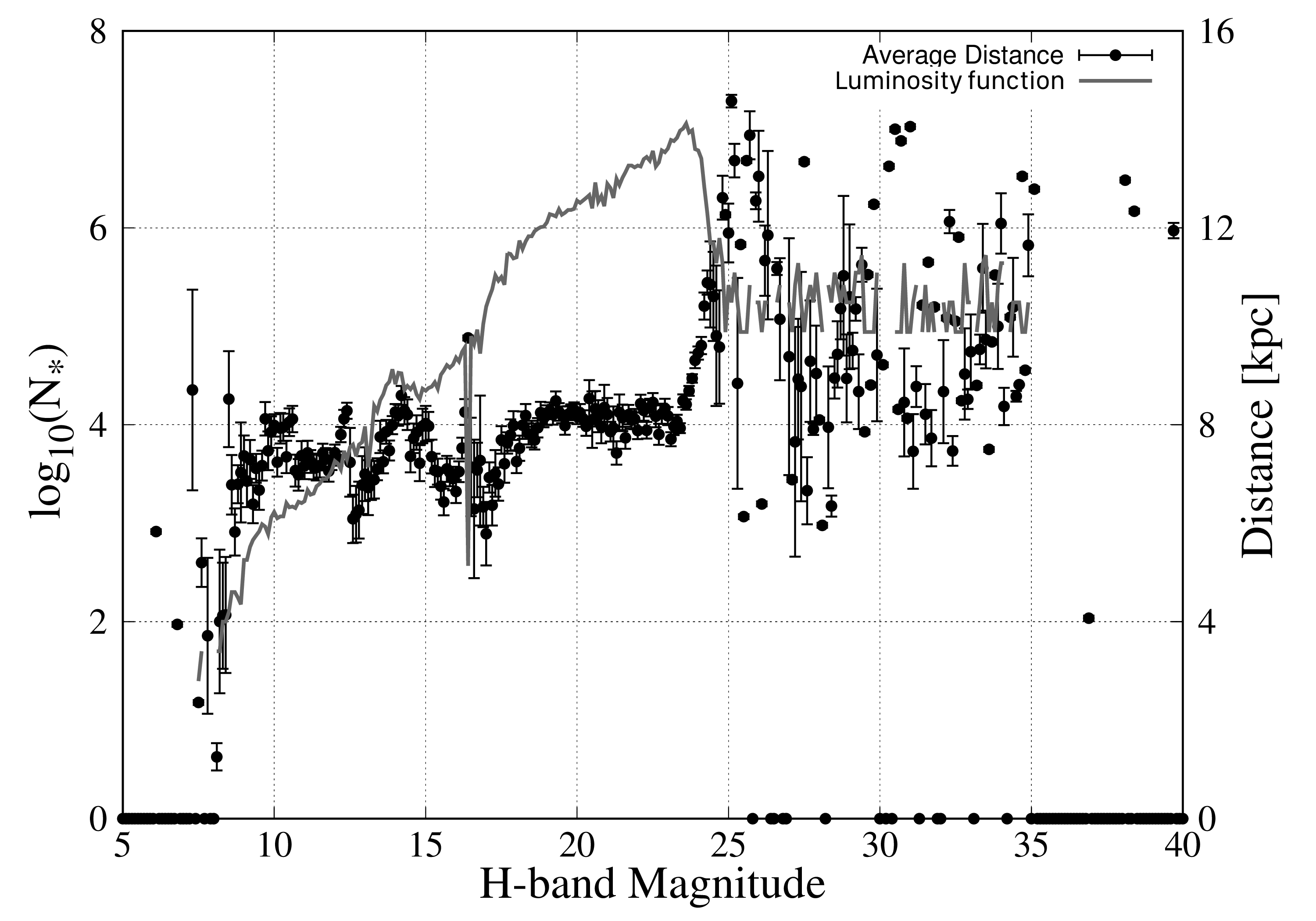}
\vspace{-0.2in}
\caption{The $H$-band luminosity function (grey line, left y-axis) and average distance (black points, right y-axis), as computed by the \bes{} model, in 0.1 mag bins. The error bars represent the maximum-minimum range of stellar distances within each magnitude bin.}
\label{fig:Ndist}
\end{figure}

Figure \ref{fig:td} shows the distribution of the fraction of detectable events in Einstein timescale and lens distance combination.  Figure \ref{fig:Ndist} is a supportive plot, showing the stellar distance and the luminosity function from the whole \bes{} catalogue. Note that both source and lens distances were taken from the same \bes{} datasets. For lenses up to 6 kpc from the Sun, the lens properties tend to be taken from the bright-stellar data ($H<9$) whose population is very low. Moreover, the disc-disc microlensing dominated the event weights because the small relative proper motion of the disc-disc microlensing offers longer events than the disc-bulge microlensing. So the fraction distribution spreads more on the long timescale regime. From 6 to 9 kpc, the timescale range spreads and the most likely distance for a detectable FFP lens is at $\sim$7 kpc for all FFP lens masses. Stars between 6-9 kpc dominate the objects simulated in our \bes{} catalogue, and occupy a wide range of $H$-band magnitudes, hence we expect a lens will typically form part of this dense population, but will normally be closer to us than the Galactic Centre (8kpc).  

Sources with $H<$23 tended to be located closer to the lens (both would be in bulge). They tend to exhibit finite source effects. The motion of bulge stars varied so that events occur with different timescales. Stars beyond 9 kpc are also included in our simulation. It is obvious from Figure \ref{fig:Ndist} that these events only had fainter sources ($H>$24). Since we considered the background noise due to unresolved stars and the extinction decreases the apparent population, these fainter sources were quite difficult to observe, and therefore, the likelihood diminishes rapidly beyond 9 kpc. According to Figure \ref{fig:Nmass}, stars with $H>$26 are more massive than stars $H\sim24$, and their mean mass covers the canonical white dwarf mass. Most of the population is white dwarfs, and a few events are detectable only with a Jupiter-mass lens (Figure \ref{fig:hs-hr}).

\begin{figure*}%[!htbp]
{\centering
\includegraphics[width=7in]{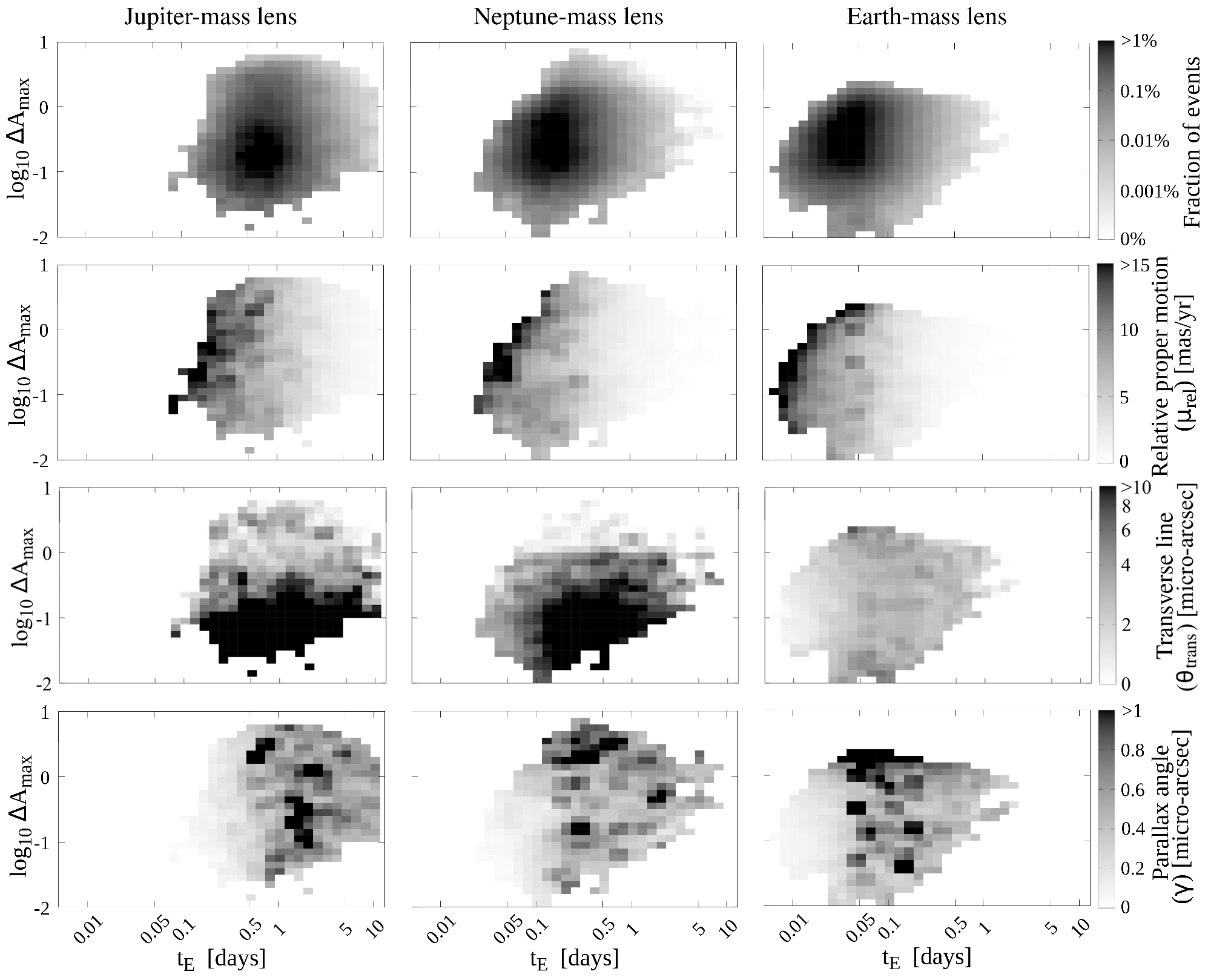}
}
\caption{Fraction of detectable events and event parameter distribution maps for the Einstein timescale and maximum differential amplitude of parallax light curves ($\Delta A_{max}$) with the bin of $\Delta t_E=0.1$ day in logarithmic calibration times $log_{10}\Delta A_{max}=0.1$.  The rows are the distributions of the fraction of detectable events, median relative proper motion, median threshold lens radii, and median parallactic angle from top to bottom. These median values are taken based on the median rate-weight in each bin.}
\label{fig:ta}
\end{figure*}

\subsection{Parallax event characteristics}	\label{subsec:res_chara}

Figure \ref{fig:ta} shows the binned distribution of some event parameters with differential amplitude and Einstein timescale. The fraction of detectable events distribution (top-row) becomes an indicator to read the relative proper motion map (2nd-row), the transverse line of the projected source in the lens frame map (3rd-row), and the parallactic angle map (bottom-row). Note that parallactic angle is time-dependent throughout an event so that we picked the maximum angle of every event (corresponding to the peak of the residual light curve) and took the mean of each bin to show the distribution. These maps are for the \euclid{}-\wfa{} combination. The figures for the \euclid{}-\lsst{} combination show similar distributions.

From the top panels of Figure \ref{fig:ta}, we can identify the most-likely combination of $(log_{10}\Delta A_{max}, t_E)$ to be at $\sim$(-0.8, 0.6), $\sim$(-0.6, 0.2) and $\sim$(-0.6, 0.03) for Jupiter-mass, Neptune-mass and Earth-mass, respectively. We can interpret the remaining panels of Figure \ref{fig:ta} through this probability distribution.

The distributions of relative proper motion (2nd-row) show a gradation along the timescale axis for all FFP lens masses. The short timescale and high differential amplitude regime corresponds to the largest proper motion. This is plausible because large relative proper motion leads to quicker events. Besides, the most likely combination of $(log_{10}(\Delta A_{max}), t_E)$ from the top panels typically yields $\mu_{rel}\sim$7.5 mas yr$^{-1}$ for all FFP lens masses. Since we used stellar data for lens properties, replacing only their mass by planetary masses, the most likely relative proper motion is similar to the mean proper motion of disc stars in our model. FFPs possibly have higher velocity as a result of ejection from their host stars and of swing-by acceleration by encounters. In that case, $\mu_{rel}$ becomes higher and yields shorter $t_E$ though we did not model it.

We calculated the mean transverse line, which is defined as the angular distance of the projected source path in the lens frame and plotted the distribution (3rd row). The Jupiter-mass and Neptune-mass FFP lenses show large size is correlated with low differential amplitude. If the radius of the Einstein ring is large compared to the projected telescope baseline, both telescopes will experience similar amplification and the light curves will not be differentiable from each other. This tendency can be seen for Earth-mass FFP events to some extent but not as extreme as massive FFPs since an Earth-mass FFP lens is small enough not to let telescopes to induce similar light curves. Like the distributions of relative proper motion, the most likely combination of $(log_{10}(\Delta A_{max}), t_E)$ from the top panels provides the transverse line $\theta_{trans}\sim8\pm2$ micro-arcsec for Jupiter-mass and Neptune-mass, and $\sim3\pm1$ mas for Earth-mass FFP. The reason is the same above; a massive lens requires a smaller threshold impact parameter to identify the differential light curve so that the source population converges to an effective transverse line. This condition indicates that a Jupiter-mass lens has the potential to observe fainter sources with which an Earth-mass lens is influenced by the finite source effect and cut-off. Thus, our criteria for detectable parallax give weight to both massive and less-massive FFPs. For the canonical Einstein radii ($\sim$0.03, $\sim$0.007, and $\sim$0.002 milli-arcsec with $D_S=8$ and $D_L=4$ for Jupiter-mass, Neptune-mass, and Earth-mass lenses, respectively), the transverse line of 8 micro-arcsec is $\sim0.3\theta_E$ for Jupiter-mass, $\sim\theta_E$ for Neptune-mass and $\sim1.5\theta_E$ for Earth-mass. Since we set the maximum impact parameter of three telescopes as $u_{max}=3$ or equivalent in the finite source case, the transverse line of $\geq\theta_E$ with any $u_0>0$ is fairly possible. However, the Earth-mass FFP events are most likely cut-off due to a finite source effect. From this point, the Neptune-mass lens is the most suitable size for the telescope separations among the three FFP masses in our model.

The distributions of the parallactic angle (bottom-row) show two regimes: the small parallactic angle regime at short timescales and the large parallactic angle spots scattered across longer timescales. The small-angle regime corresponds to our parallax detectability limit, i.e., that the event is simultaneously observable by both telescopes at least for 1 hour under the proper SAA, day/night timing, and weather. Especially for low-mass lens events, a large proper motion reduces the chance of ``simultaneous'' observation so that a similar line-of-sight between two observers is preferred to satisfy our detectability limit. The large parallactic angle spots are likely an artefact due to plotting the mean of each bin. We assume that the maximum differential amplitude corresponds to the maximum parallactic angle and take their maximum values individually. This assumption is plausible since the telescope separation is quite smaller than the source and lens distances; however in a spatial model of our simulation, this assumption is not always the truth. The model treats the direction of their line-of-sights in vector, and the parallax angle is found in the way of simulation whilst it would be found from the differential light curve at real observations. For the Earth-mass FFP lens, the large parallactic angle spots seem to concentrate on the high differential amplitude regime. A large differential amplitude requires relatively wider telescope separation or large difference in time of maximum amplification. In either case, a very small impact parameter is necessary for Earth-mass FFP lens to offer $log_{10}(\Delta A_{max})$>0.2 satisfying the detectable duration limit of $T>$1 hour. The most likely combination of $(log_{10}\Delta A_{max}, t_E)$ from the top panels also has similar parallactic angles for all FFP lens masses: $\gamma$$\sim$0.36$\pm$0.02 micro-arcsec. As for the threshold lens radii, this is because the requirement of threshold impact parameter and the Einstein radius concentrated parallactic angles towards a specific value across all FFP lens masses.

\begin{figure}	
\hspace{-0.1in}\includegraphics[width=3.5in]{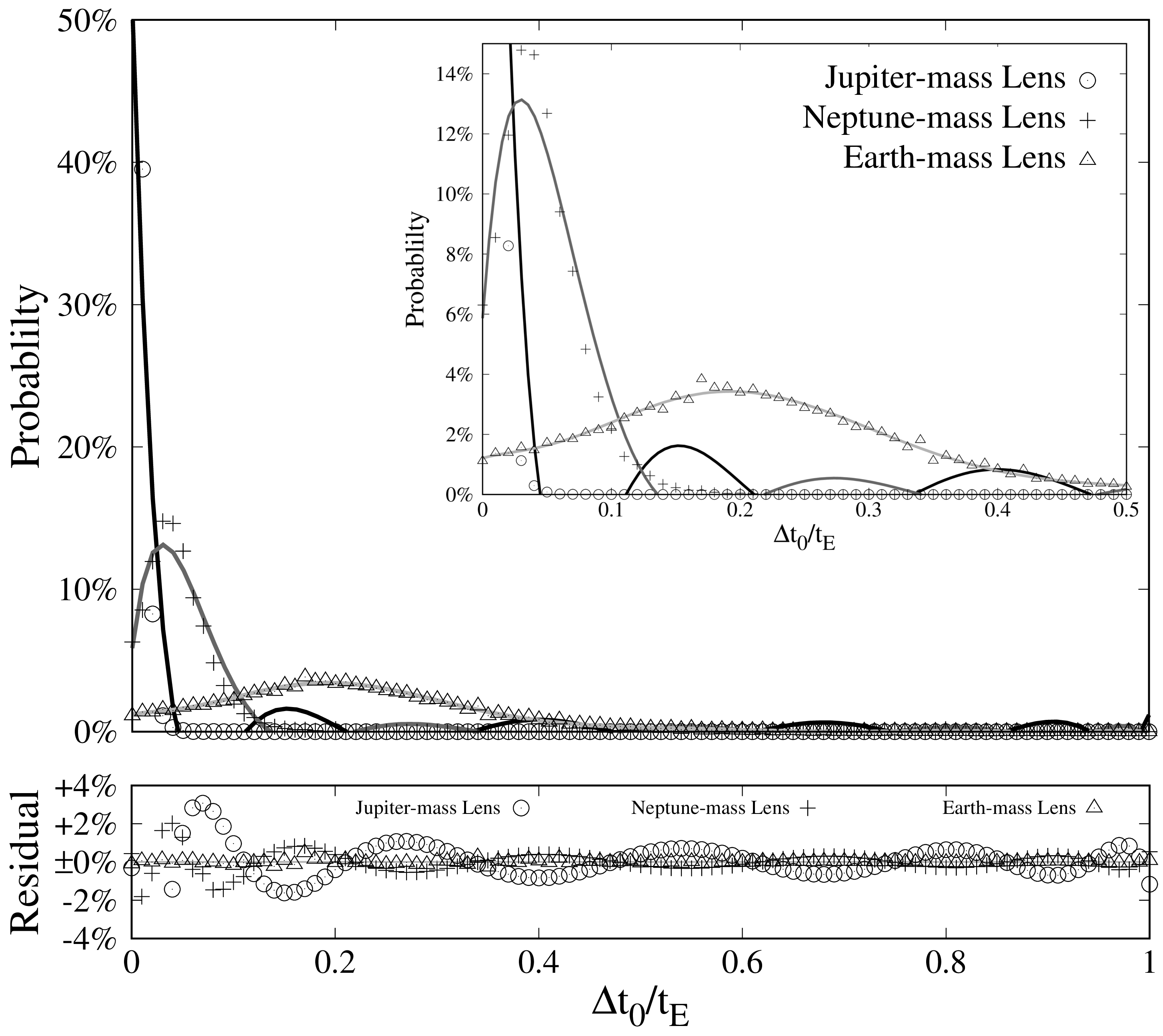}	
\caption{Probability distribution of difference in time of maximum amplification between light curves detected by the \euclid{}-\wfa{} combination. The \euclid{}-\lsst{} combination provided a very similar distribution. The black, dark grey and light grey fitting curves are 10th polynomial fits. Residuals from these fits are shown in the bottom panel. The insert shows a zoom of the bottom-left corner.}
\label{fig:time}
\end{figure}

Figure \ref{fig:time} shows the probability distribution of difference in time of maximum amplification. Lower FFP masses generate longer times between the two light curve peaks. This is because the projected telescope separation becomes larger with respect to the threshold lens radii. For the Neptune-mass and Earth-mass FFP lenses, the highest probability is at $\Delta t_0/t_E$$\sim$0.05 and $\Delta t_0/t_E$$\sim$0.19, respectively. For the most-likely event timescale, the highest probability corresponds to a time difference of  $\sim$33 min for Neptune-mass lens and $\sim$22 min for Earth-mass lens. These values are much larger than the maximum time-gap due to the observer separation: taking $\leq$6 sec for radiation travel between Earth and L2. Besides, it is larger than or close to the cadence of our assumed telescopes (15-20 min). The time gap is even possible to identify through real observation.

\subsection{Event rate}	\label{subsec:res_er}
The event rate is summarised in Table \ref{tab:eventrate}. The yearly rate covers  two 30-day-seasons around the equinoxes, as determined by the maximum SAA of \euclid{} (\$\ref{subsec:tele-kine}). For the \euclid{}-\wfa{} combination, we simulated both geosynchronous orbit and Halo orbit at L2 cases of \wfa{}. The parallax probability is calculated from the weighted rate of the event (Eq.(\ref{eq:wrate})) where the threshold impact parameter ($u_t$) depends on the telescope configurations. Therefore, there might be a gap of parallax probability between the telescopes even though they are detecting the same events. It is reasonable to accept the smaller probability between the combination telescopes rather than optimising the large number. For the \euclid{}-\lsst{} combination, the parallax detectability of our model considers the day-night position of \lsst{}. Our assumption of 7.5 hours of performance per night corresponds to the zenith angle of $\sim56^{\circ}$. Hence, the population of clear nights (65.89\%) is only applied to the final result.

For the \euclid{}-\wfa{} combination, the Halo orbit at L2 case provided a lower probability than the geosynchronous orbit case. This arises mostly from the observer separation in our simulation. On average, the geosynchronous orbit case provided the observer separation $\sim$1.6$\times10^6$ km whilst the Halo orbit at L2 case provided $\sim$0.9$\times10^6$ km. The narrower separation reduces the light curve gap. Moreover, the variation of the relative line-of-sight is greater for the Halo orbit at L2 case than the geosynchronous orbit case because of its motion apart from the reference observer on the Earth. Whilst the geosynchronous orbit case keeps the parallactic angle based on the SAA, the Halo orbit at L2 is more likely to yield a small parallactic angle. We confirmed that the distribution of sample event properties (i.e. lens distance, lens size, relative proper motion and threshold impact parameter) did not show a clear difference between these cases. Therefore, we can simply say that the variation of phase positions resulted in a smaller parallax event rate for the Halo orbit at L2 case than the geosynchronous orbit case. In both orbital cases, the parallax event rate derived from the \wfa{} configuration is smaller than that of \euclid{} configuration. It indicates that more events are detected only by \wfa{} in our simulation and is understandable because the \wfa{} sensitivity yields the fainter zero-point magnitude and allows more faint sources which population is relatively large.

\begin{table}
 \centering
  \caption{FFP parallax event rate $\tilde{\Gamma}_{parallax}$ [events year$^{-1}$ deg$^{-2}$] targeting at $(l, b) = (1^{\circ}, -1.^{\circ}75)$. The parallax probability ($P_{parallax}$) is given in parentheses. Note that $P_{parallax}$ is the summation of the weighted rate of detectable parallax events overall events (detected either in parallax or solely.) in a year. The same values of parallax probability may result in different parallax event rate because of the different FFP event rate of the solo-observation for each telescope (Eq.(\ref{eq:prlx_er})), and the year$^{-1}$ unit of the parallax event rate means ``per yearly co-operation period'' (i.e. two 30-days operation of \euclid{}). EW stands for the \euclid{}-\wfa{} combination whilst EL stands for the \euclid{}-\lsst{} combination. Geo and L2 show the orbital options of \wfa{}.}
  \label{tab:eventrate}
  \begin{tabular}{ll|lll}
  \hline

\multicolumn{2}{c|}{Lens mass} & Jupiter & Neptune & Earth \\ \hline
\multirow{2}{*}{EW-Geo} & \euclid{} & 80.3 (3.9\%) & 34.1 (7.2\%) & 11.4 (10.0\%) \\
			       & \wfa{} &  52.0 (2.6\%) & 19.0 (4.0\%) & 4.8 (4.3\%) \\ \hline
\multirow{2}{*}{EW-L2} & \euclid{} & 45.0 (2.2\%) & 20.0 (4.2\%) & 6.7 (5.8\%) \\
			       & \wfa{} & 30.7 (1.5\%) & 13.3 (2.8\%) & 3.9 (3.4\%) \\ \hline
\multirow{2}{*}{EL} & \euclid{} & 34.5 (2.6\%) & 8.9 (2.8\%) & 0.5 (0.7\%) \\
			     & \lsst{}  & 47.5 (3.9\%) & 14.1 (5.1\%) & 1.0 (1.5\%) \\	\hline
\end{tabular}
\end{table}

The \euclid{}-\lsst{} combination yields less FFP parallax detection than the \euclid{}-\wfa{} combination with the Halo orbit. This event rate is based on our fine-weather assumption of 65.89\% and an average operation of 7.5 hours/night. Compared with the geosynchronous orbit case (the telescope separation is similar to the \euclid{}-\lsst{} combination), the parallax event rate becomes smaller because the event detectability is lead by less-sensitive \lsst{}. The event rate derived from the \euclid{} configuration is smaller than that from the \lsst{} configuration, unlike the \euclid{}-\wfa{} combination. The reason is the same as the \euclid{}-\wfa{} combination that there are more events observed only by \euclid{} because of the sensitivity difference in our simulation.

The combination of \wfa{} and \lsst{} was skipped in our simulation. The zero-point magnitude difference would yield the sensitivity difference and result in different detectability, but as long as \lsst{} is the less-sensitive ground-based survey, the \lsst{} sensitivity determines the detectability like the \euclid{}-\lsst{} combination. Besides, the geosynchronous orbit case would provide an observer separation that is too short to successfully and effectively observe parallax. For the Halo orbit case, the wider SAA of \wfa{} would provide a larger event rate than the \euclid{}-\lsst{} combination, and we can simply multiply the event rate by the SAA coverage ratio to gain the \wfa{}-\lsst{} combination event rate since our random selection of Earth's positioning angle (hence, L2 position) is equally distributed within the range.

\citet{Zhu2015} and \citet{Zhu2016} considered parallax observations by ground-based and space-based surveys. They tested the combination of OGLE-{\it Spitzer} and of KMTNet-\wfa{} where the Halo orbit case of \wfa{} was assumed. In both combinations, they suggested the importance of telescope separation to observe FFPs as we have discussed in relation to our results. Besides, \lsst{} is expected to have higher sensitivity than other ground-based telescopes currently operating. Although \citet{Zhu2015} suggested the reinforcement of sensitivity by a ``combination'' of ground-based and space-based telescopes, the individual sensitivity is essential to observe FFP events effectively.

\begin{table}
 \centering
  \caption{The accuracy of mass estimation for the 30,000 simulated events towards $(l, b) = (1^{\circ}, -1.^{\circ}75)$. The percentile likelihood indicates the ratio of events for which the uncertainty ($\epsilon$) in the estimated mass ($M_{FFP}$) successfully covers the given mass ($M_{given}$). EW stands for the \euclid{}-\wfa{} combination whilst EL stands for the \euclid{}-\lsst{} combination. Geo and L2 are the orbital options of \wfa{}.}
  \label{tab:uncM}
  \begin{tabular}{ll|lll}
  \hline
  \multicolumn{2}{c|}{$M_{FFP}\pm\epsilon \supset M_{given}$} & Jupiter & Neptune & Earth \\ \hline

\multirow{2}{*}{EW} & Geo         & 23.7\%  & 27.5\%   & 34.6\% \\ 
			 & L2           & 20.1\%  & 24.7\%  & 33.1\% \\
                   EL  &                & 18.4\%  & 21.0\%   & 23.6\% \\	\hline

\end{tabular}
\end{table}

\begin{figure}%[!htbp]	
\hspace{-0.1in}\includegraphics[width=3.5in]{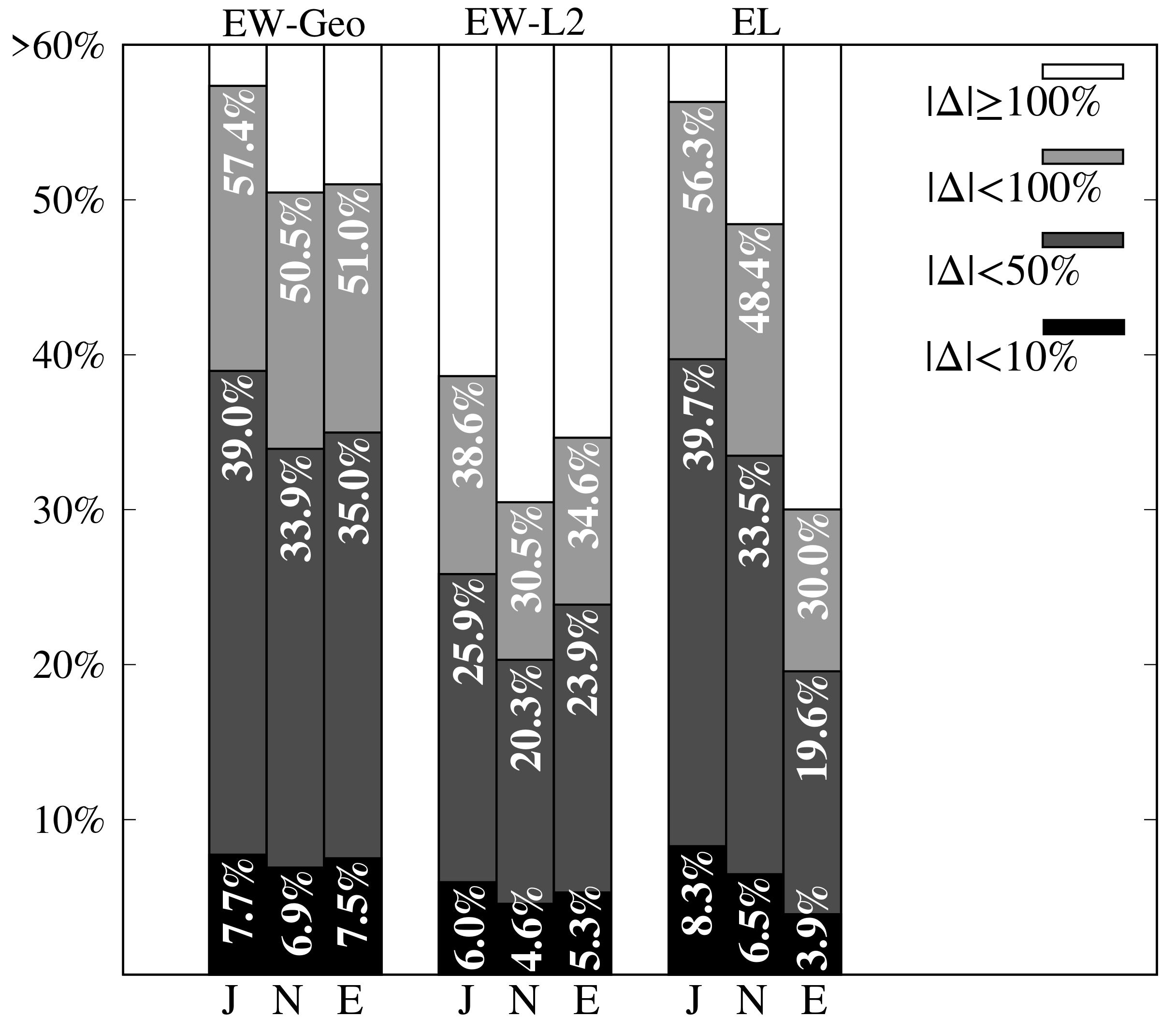}	
\caption{Likelihood of obtaining a lens mass to with a fraction $|\Delta|$ of the true lens mass among our simulated events. The discrepancy was calculated as $|Delta|$=$|M_{FFP}-M_{input}|$/$M_{input}$. EW stands for the \euclid{}-\wfa{} combination whilst EL stands for the \euclid{}-\lsst{} combination. Geo and L2 are the orbital options of \wfa{}. The Jupiter-mass, Neptune-mass, and Earth-mass FFP distributions are labelled by their initials.}
\label{fig:disMbar}
\end{figure}

\subsection{Accuracy of mass estimation from parallax light curves}	\label{subsec:res_mass}
The mass estimation of the lens will theoretically be more precise using parallax data than a single observation. For every detectable event in our simulation, the residual light curve of parallax observation was generated. $u_1(t)$, and $u_2(t)$ are derived from the light curve.  Instead of taking all $u$ for every $t$, we took $u$ at the light curve peak of each observer (see Figure \ref{fig:prlx-geo} where is expressed as $\Delta A(t_{0,1})$ and $\Delta A(t_{0,2})$); hence we had $u_1(t_{0,1})$, $u_2(t_{0,1})$, $u_1(t_{0,2})$, and $u_1(t_{0,2})$. The relative proper motion was assumed to be $\mu_{rel}\sim7.5\pm1.5$ mas yr$^{-1}$, which we derived from the most likely values of ($log_{10}(\Delta A_{max}), t_E$; Figure \ref{fig:ta}). Then $\theta_E$ was calculated. The telescope separation for each combination is averaged as $D_T\sim1.6\times10^6$ km for the \euclid{}-\lsst{} combination and $\sim0.9\times10^6$ km for the \euclid{}-\wfa{} combination, and the fluctuation by the orbital motion is contained as the uncertainty of it. Using Eq.(\ref{eq:prlx_angle}) and these parameters from the light curve, $\gamma$ was calculated.  Thus, we had $\gamma(t_{0,1})$ and $\gamma(t_{0,2})$ and moved to Eq.(\ref{eq:prlx_mass}) to calculate the lens mass and averaged them.

To consider the accuracy of our lens mass estimation, we calculate the uncertainty ($\epsilon$) from the parameter errors of $u$, $\mu_{rel}$, and $D_T$ and the discrepancy of the estimated mass from the input mass ($|Delta|$=$|M_{FFP}-M_{input}|$/$M_{input}$). If $M_{FFP}\pm\epsilon$ does not cover the input mass, it is obvious that the uncertainty is underestimated. However, even though $M_{FFP}\pm\epsilon$ covers the input mass, the very large $\epsilon$ case is not acceptable from the view of the estimation accuracy. Table \ref{tab:uncM} summarises the percentage likelihood among our 30,000 events that $M_{FFP}\pm\epsilon$ covers the input mass. It indicates that our parameter errors are underestimated in the majority case. It is because we assumed a certain value of relative proper motion (7.5  mas yr$^{-1}$) with an error value that was derived from the \bes{} data. Considering the distribution of detectable parallax events, the error is flexible in dependance on the differential maximum-amplitude.  Figure \ref{fig:disMbar} visualises the percentage likelihood of the discrepancy among 30,000 events. For example, $|\Delta|$<100\% means the estimated mass was up to twice as large as the input mass. Since $|\Delta|$<50\% regime is less than 50\% for all FFP masses and telescope combinations, we can regard that our calculation likely overestimates the FFP mass. One cause was that the parallactic angle was underestimated in most cases because we derived it from the differential light curves, not from the spatial components in our model.  The other cause was that we theoretically estimated the impact parameter to derive the parallactic angle from the light curve using Eq.(\ref{eq:au}) assuming a point source. Thus, the parallactic angle variation was wider than what we generalised through our mass estimation process, and our calculation method biased to make the least parallactic angle.

\begin{table}
 \centering
  \caption{The mean estimated mass and the standard deviation for the representative data of $|\Delta|$<10\%, <50\%, and <100\%. These data can be regarded as showing the Gaussian distribution for each discrepancy range. The values are in units of each input mass.}
  \label{tab:gaussM}
  \hspace*{-0.3in}
  \begin{tabular}{|ll|cc|cc|cc|}
  \hline
   & &  \multicolumn{2}{c}{Jupiter} & \multicolumn{2}{c}{Neptune} & \multicolumn{2}{c}{Earth} \\ 
  				        &            & Mean & SD & Mean & SD & Mean & SD \\ \hline
  \multirow{3}{*}{EW-Geo} & $|\Delta|$<10\% & 1.00 & 0.06 & 1.00 & 0.06 & 1.00 & 0.06 \\
  					& $|\Delta|$<50\% & 0.95 & 0.28 & 0.96 & 0.28 & 0.96 & 0.28 \\
  					& $|\Delta|$<100\% & 1.04 & 0.46 & 1.04 & 0.46 & 1.02 & 0.46 \\ \hline
  \multirow{3}{*}{EW-L2}    & $|\Delta|$<10\% & 1.00 & 0.06 & 1.00 & 0.06 & 1.00 & 0.06 \\
  					& $|\Delta|$<50\% & 1.03 & 0.27 & 1.03 & 0.27 & 1.01 & 0.27 \\
  					& $|\Delta|$<100\% & 1.21 & 0.43 & 1.21 & 0.43 & 1.18 & 0.43 \\ \hline
  \multirow{3}{*}{EL}          & $|\Delta|$<10\% & 1.00 & 0.06 & 1.00 & 0.06 & 1.00 & 0.06 \\
  					& $|\Delta|$<50\% & 0.95 & 0.28 & 0.93 & 0.28 & 1.00 & 0.28 \\
  					& $|\Delta|$<100\% & 1.05 & 0.44 & 1.01 & 0.45 & 1.15 & 0.46 \\ \hline
\end{tabular}
\end{table}

Table \ref{tab:gaussM} shows the Gaussian means and standard deviations derived from the data of each discrepancy: <10\%, <50\%, and <100\%. As we mentioned in the last paragraph, the mean estimated masses tend to become equal to or larger than the input mass because of the underestimation of the parallactic angle. However, it is not true for the Neptune-mass case in the \euclid{}-\wfa{} combination with the geosynchronous orbit and the \euclid{}-\lsst{} combination. One possible reason is because of the interaction between Einstein radius and the Einstein parallax (i.e. the telescope separation). In \S\ref{subsec:res_chara}, we confirmed that the Neptune-mass lens is the most suitable to our telescope combinations because of the effectiveness of the transverse line for identify the differential light curve. In other words, the upper limit of the minimum impact parameter, which satisfies our parallax detectability limit, was maximised for the Neptune-mass lens. Thus, Table \ref{tab:gaussM} indicates that the telescope separation of $1.6\times10^6$ km allowed relatively larger $u_0$ and underestimated the Einstein radius, which we derived from the transverse line and the detected duration reading from the light curves.

The discrepancy of estimated FFP lens mass correlates with the accuracy of Einstein timescale ($t_E$) and Einstein parallax ($\pi_E$) estimated from light curves. Note that Einstein parallax was calculated from the reciprocal of observer separation corresponding to the full lens size. We confirmed that the accuracy of Einstein timescale estimation was quite acceptable; 74.5\% of events had Einstein timescales reproduced to within 10\% of the input timescale, and almost all of the rest stayed within 50\% discrepancy. On the contrary, Einstein parallax estimation was as inaccurate as the FFP mass estimation. Hence, we need to improve the Einstein parallax estimation from light curves. In our simulation, the 3D positioning model was used, and the measurement of observer separation was done in vectorial 3D space. However, photometric light curves were drawn with the scalar value of impact parameters since we never know the vectorial impact parameters on the sky during real observations, since we never know the rotation angle that the transect the source star makes behind the FFP lens.

We computed the distribution of the angular difference between two vectorial impact parameters observed by two telescopes. Here we define the angular difference as $\alpha_u$  that two vectorial impact parameters $\vec{u_1}$ and $\vec{u_2}$ form between. 

The angle $\alpha_u$ is not evenly distributed. The most likely case was the opposite vectorial direction (i.e. $\alpha_u\sim\pi$), but the case was just 11-29\% of detectable parallax events depending on the observer combinations in our simulation, which allowed $\alpha_u$ to take from $-\pi/2$ to $\pi/2$. Compared to the event rate values in Table \ref{tab:eventrate}, we also found that the percentile likelihood of $\alpha_u\sim\pi$ decreased for the higher event rate. One reason for this tendency was that the relative positions between two telescopes varied more for the \euclid{}-\wfa{} combination than for the \euclid{}-\lsst{} combination. Another reason was that the \euclid{}-\lsst{} combination required a larger amplitude difference between two light curves due to the lower sensitivity of ground-based partner (\lsst{}).

Thus, the Einstein parallax estimation using a scalar value of impact parameters was not enough accurate in the commonest scenario, and this issue affected our mass estimation accuracy as we discussed above. We also tried to compute a distribution of parallax time gap (i.e. difference of the light curve peak time between two observers divided by the Einstein timescale; $\Delta t_0/t_E$) for every $\alpha_u$. The range of possible parallax time gap showed a convex curve for $-\pi/2\leq\alpha_u<\pi/2$, but the distribution peak depends on the telescope combinations and FFP masses. Thus, the parallax time-gap was not sufficient evidence to identify the angle between two impact parameters. Besides, the error in the time of peak magnification ($\epsilon(t_0)$) should be smaller than the parallax time gap otherwise the fraction error in Einstein parallax exceeds unity.

% E : 2022- for 6.25 years, 0.54 deg2 FoV
% W : 2024- for 5 years, 0.281 deg2 FoV
% L : 2020-, 9.6 deg2 FoV

\section{Discussion}	\label{sec:disc}
For the \euclid{}-\wfa{} combination, the phase difference influences the differential light curve more on the low-mass FFP lenses. In our model, we assumed the phase difference between the \euclid{} and \wfa{} position 90 degree, which corresponds to $\sim0.9\times10^6$ km separation. If it was the maximum phase difference of 180 degree, the separation would be $\sim1.2\times10^6$ km separation. The additional run of the L2 orbit case with a 180-degree phase difference simulation showed an increase of the parallax event rate from the 90-degree phase difference case but not as large as the event rate of the geosynchronous orbit case (the telescope separation is $\sim1.6\times10^6$ km). Hence, the phase difference between \euclid{} and \wfa{} is one of the important issues to effectively yield the parallax detection.

We assumed the FFP population is 1 per star, and we can numerically convert our result to the other population cases. For instance, there are several opinions of the Jupiter-mass FFP population per MS star: $\sim$1.8 \citep{Sumi2011}, 1.4 \citep{Clanton2016}, and <0.25 \citep{Mroz2017} though the detailed conditions and assumptions are different. Therough the microlensing event simulation ($\S$\ref{sec:micro}), the optical depth is proposional to the lens population. This means that we can define the conversion formula of the FFP event rate for the different FFP population as
\begin{equation}     \label{eq:convertion}
\tilde{\Gamma}_{FFP}^{new} = \tilde{\Gamma}_{FFP} \times f_{conv} \hspace{0.2in} {\rm where} \hspace{0.2in} f_{conv} \propto P_{FFP}, 
\end{equation}
where $\tilde{\Gamma}_{FFP}$ is the actual FFP microlensing event rate per year which value we used in the simulation is in Table \ref{tab:ffp-ev} and $P_{FFP}$ is a population ratio of FFPs per star. According to Figure \ref{fig:hs-hr}, the MS stars are $H$>16.5 which roughly corresponds to the boundary of catalogue B and C. In our simulation, the 99.6\% of source and lens data refer from the catalogue C and D due to the population. Therefore, we can approximate the conversion coefficient as $f_{conv}\sim$1.8, $\sim$1.4 and $\sim$0.25 for \citeauthor{Sumi2011}'s, \citeauthor{Clanton2016}'s and \citeauthor{Mroz2017}'s FFP population, respectively. Since 99.6\% of our source stars are MS stars, we can simply multiply the rates in Table \ref{tab:eventrate} by these factors (the associated probabilities do not change). As a result, \citeauthor{Sumi2011}'s population predicts that the \euclid{}-\wfa{} combination observe 55 Jupiter-mass FFP for two 30-day periods per year in parallax. The \citeauthor{Clanton2016}'s population is 43 FFPs, and \citeauthor{Mroz2017}' population is <8 FFPs.

We numerically modelled those noise sources that could easily be quantified. In the real observation, there is the effect of the planetesimals and asteroids in the asteroid belt, Kuiper belt, and Oort cloud, and the flux interference by the stellar flares, transits of the source system, and binary-source events. The effect from asteroids is not negligible if the orbits scratch the line-of-sight of the event, and the same thing around the source star offers the flux noise \citep{Trilling2005, Usui2012, Matthews2014, Wong2017, Whidden2019}. The stellar flares provide a sudden magnitude increase, and the followup observation with high cadence will be required to identify either a stellar flare or a short microlensing event from the light curve \citep{Balona2016}. The transit of the planetary system requires the reduction of the phase from the light curve to identify the one-time microlensing event\citep{Hidalgo2018}. The binary source makes the light curve more complex than the single-source events we assumed \citep{Kong2011}. Thus, the determination of short events from the light curve variation becomes difficult, and the event rate will be less than we derived here.

\section{Conclusion}	\label{sec:conc}
We have simulated the parallax observations of FFP events in a 3D model, targeted towards $(l, b)=(1^{\circ},-1.^{\circ}75)$. \euclid{} was taken as the main telescope and two different partners were applied; \wfa{} and \lsst{}. We estimate that the \euclid{}-\wfa{} combination will result in 3.9 Earth-mass and 30.7 Jupiter-mass FFP microlensing events will be detectable with sufficient sensitivity to determin their parallax angle during two 30-day-periods of \euclid{} operation per yearly co-operation period per square degree. From the latest operation plan for \euclid{} and \wfa{}, we may expect that the chance of simultaneous parallax observation will be $\sim$2.5 years (or 5$\times$30-day operations) with 0.28 deg$^2$ field-of-view (FoV). This results in 2.7 Earth-mass FFPs and 21.5 Jupiter-mass FFPs that will be observed with simultaneous observations and measurable parallax. On the other hand, the \euclid{}-\lsst{} combination resulted in less event rate due to our optimisation of weather and night, and about 0.5 Earth-mass and 34.5 Jupiter-mass FFPs can be found per year per square degree. Unlike the \euclid{}-\wfa{} combination, \lsst{} is capable to cover the expected \euclid{} operation period of $\sim$6 years and 0.54 deg$^2$ FoV. As results in 1.8 Earth-mass FFPs and 112 Jupiter-mass FFPs will be observed in simultaneous parallax. As we mentioned at the introduction of telescopes applied to the simulation ($\S$\ref{subsec:tele-kine}), the exoplanet research campaign using microlensing observation is planned in the \euclid{} and \wfa{} missions but not in the \lsst{} survey. Out result at least shows the potential of parallel ground-based observing microlensing with the collaboration with upcoming space-based surveys.

The mass estimation from parallax light curves still has some problems with accuracy. In our calculation, the estimated mass is only accurate to <20\% with the uncertainty of $\Delta$<0.5$M_{FFP}$ for both the \euclid{}-\wfa{} combination in the \wfa{} Halo orbit case and the \euclid{}-\lsst{} combination. The vectorial impact parameters are the main source of uncertainty in the mass estimation in our simulation, and the additional approaches we attempted could not over come this. Improved methods of estimating event configuration angle will be required for improved mass estimation.

In our simulation, we considered two orbital options for \wfa{}; the geosynchronous orbit and the Halo orbit at L2. Our simulation made clear that the telescope separation of both cases is useful for a microlensing parallax observation from Earth-mass to Jupiter-mass FFPs. The  \euclid{}-\wfa{} combination with the geosynchronous orbit resulted in a slightly larger event rate than with the Halo orbit at L2 because of the difference in the observer separation. However, further research about noise detection and reduction is necessary, especially for low-mass FFPs.  Besides, we did not consider the variation in \euclid{} and \wfa{} trajectories. The probability of parallax detection we have suggested in this paper will be a criterion for further research and future observations. The simultaneous parallax observation is expected to explore the study of exoplanets, including FFPs, in the next decades.

\section{Acknowledgement}	\label{sec:ackn}
We would like to thank Dr. Kerins for his expert advice to begin this research and Dr. Robin and her team of \bes{} Galactic model to offer us an important source for the simulation.

%%%%%%%%%%%%%%%%%%%%%%%%%%%%%%%%%%%%%%%%%%%%%%%%%%

%%%%%%%%%%%%%%%%%%%% REFERENCES %%%%%%%%%%%%%%%%%%

% The best way to enter references is to use BibTeX:

\newpage
\bibliographystyle{mnras}
\bibliography{ref}

%%%%%%%%%%%%%%%%%%%%%%%%%%%%%%%%%%%%%%%%%%%%%%%%%%

% Don't change these lines
\bsp	% typesetting comment
\label{lastpage}
\end{document}